\documentclass[a4paper,11pt]{article}
\usepackage{natbib}
\usepackage[colorlinks,citecolor=blue,urlcolor=blue]{hyperref}
\usepackage{epsf}
\usepackage{epsfig}
\usepackage{multirow}
\usepackage[english]{babel}
\usepackage{graphicx}
\usepackage{geometry}
\usepackage{natbib}
\usepackage{amssymb}
\usepackage{amsmath}
\usepackage{mathtools}
\mathtoolsset{showonlyrefs=true}
\usepackage{color}
\usepackage{enumerate}
\usepackage{authblk}
\usepackage[table,xcdraw,dvipsnames]{xcolor}
\providecommand{\keywords}[1]{\textbf{\textit{Keywords---}} #1}
\usepackage{booktabs}
\usepackage{makecell}

\usepackage[font={small,it}]{caption}
\usepackage[labelformat=simple]{subcaption}
\usepackage{bm}

\newcommand{\specificthanks}[1]{\@fnsymbol{#1}}

\title{Model-based clustering for network data via a latent shrinkage position cluster model}

\author[]{Xian Yao Gwee\thanks{xian-yao.gwee@gmail.com} }
\author[]{Isobel Claire Gormley\thanks{claire.gormley@ucd.ie}}
\author[]{Michael Fop\thanks{michael.fop@ucd.ie\\[0.2cm] Isobel Claire Gormley and Michael Fop have contributed equally to this work.}}
\affil[]{School of Mathematics and Statistics,\protect \\ University College Dublin.}

\date{}

\begin{document}
\maketitle

\begin{abstract}
Low-dimensional representation and clustering of network data are tasks of great interest across various fields. Latent position models are routinely used for this purpose by assuming that each node has a location in a low-dimensional latent space, and by enabling node clustering. However, these models fall short through their inability to simultaneously determine the latent space dimension and number of clusters. Here we introduce the latent shrinkage position cluster model (LSPCM), which addresses this limitation. The LSPCM posits an infinite dimensional latent space and assumes a Bayesian nonparametric shrinkage prior on the latent positions’ variance parameters resulting in higher dimensions having increasingly smaller variances, aiding the identification of dimensions with non-negligible variance. Further, the LSPCM assumes the latent positions follow a sparse finite Gaussian mixture model, allowing for automatic inference on the number of clusters related to non-empty mixture components. As a result, the LSPCM simultaneously infers the effective dimension of the latent space and the number of clusters, eliminating the need to fit and compare multiple models. The performance of the LSPCM is assessed via simulation studies and demonstrated through application to two real Twitter network datasets from sporting and political contexts. Open source software is available to facilitate widespread use of the LSPCM.
\end{abstract}



\section{Introduction}
\label{sec:intro}
Network data are an important class of structured data where objects are represented as nodes and the relationships between these objects are represented as edges. Network data arise in various fields, including epidemiology \citep{jo_2021_a}, neuroscience \citep{yang_2020_simultaneous}, brain connectivity \citep{aliverti_2019_spatial} and sociology \citep{dangelo_2019_latent}. A key interest in the analysis of network data is the task of clustering the nodes in the network, where the aim is to group together nodes that share similar characteristics or behaviours.

While a variety of network models exist, many originate from the influential Erd\"{o}s R\'{e}nyi random graph model \citep{erds_1959_on, gilbert_1959_random}, including the widely utilised stochastic block model \citep{holland_1983_stochastic, snijders_1997_estimation} and the latent position model \citep[LPM,][]{hoff_2002_latent}. Here, we focus on the LPM which assumes each node in a network has a position in a latent $p$-dimensional space. Under the LPM, the probability of an edge between two nodes is determined by their proximity in the latent space. The latent position cluster model \citep[LPCM,][]{handcock_2007_modelbased} extended the LPM to allow for model-based clustering of nodes in a network, by assuming the latent positions are generated from a finite mixture of Gaussian distributions. Various extensions to the LPCM have been proposed, including approaches that account for random effects \citep{krivitsky_2009_representing}, that incorporate covariates through a mixture of experts framework \citep{gormley_2010_a}, that employ a variational approach to Bayesian inference \citep{saltertownshend_2013_variational}, that model longitudinal networks \citep{sewell_2017_latent}, that allow for edge clustering \citep{sewell_2020_modelbased, Pham_Sewell_2024}, and that model multiplex networks \citep{dangelo_2023_modelbased}; see \cite{rastelli_2016_properties, Sosa_Buitrago_2021, kaur_2023_latent} for comprehensive reviews. 

Although the LPCM is widely used, inferring the number of clusters and the dimensionality of the latent space remains a challenging task. In practice, the numbers of clusters and dimensions are typically treated as user-defined parameters, with the latter set as $2$ in the majority of cases to allow for simple visualisation and interpretation \citep{dangelo_2023_modelbased, liu_2023_variational}. Often, 
a set of LPCMs with different numbers of clusters and dimensions are fitted to the network data and a model selection criterion is used to select the optimal model. Many model selection criteria have been used for this purpose e.g.,  \cite{handcock_2007_modelbased} use a variant of the Bayesian information criterion (BIC) to select optimal numbers of clusters and dimensions, but highlight its lack of robustness. Other model selection criteria such as the Watanabe–Akaike information criterion (WAIC) \citep{ng_2021_modeling, sosa_2022_a}, the Akaike information criterion (AIC) and the integrated completed likelihood (ICL) \citep{sewell_2020_modelbased} have also been used. While these criteria have found some success, they can give conflicting inference on the same data and their use requires fitting a large set of models, each with a different combination of number of clusters and number of dimensions, which becomes computationally prohibitive as the set of possible models grows.

Alternative, automated strategies for inferring the numbers of clusters and dimensions from the network data have emerged. In the context of stochastic block models, \cite{yang_2020_simultaneous} utilise a frequentist framework, while \cite{passino_2020_bayesian} employ a Bayesian framework for this purpose. In the LPCM setting, automatic inference on the number of clusters has been considered in \cite{dangelo_2023_modelbased} via an infinite mixture model, while \cite{durante_2014_nonparametric} and \cite{gwee_2022_a} 
employed nonparametric shrinkage priors to address the latent space dimensionality. However, automated, simultaneous inference of both the number of clusters and the effective latent space dimension in the LPCM setting has not yet been considered. 

Here the latent shrinkage position cluster model (LSPCM) is introduced, which simultaneously infers the node clustering structure and the effective latent space dimension. To achieve clustering of the nodes, a sparse finite mixture model \citep{malsinerwalli_2014_modelbased, malsinerwalli_2017_identifying} is employed that overfits the number of components in the mixture model. The adoption of a sparse prior on the mixture weights encourages emptying of redundant components, thereby allowing inference on the number of clusters. Within each cluster, a Bayesian nonparametric truncated gamma process shrinkage prior \citep{bhattacharya_2011_sparse, gwee_2022_a} is placed on the variance of the nodes' positions in the latent space. While the latent space is assumed to have infinitely many dimensions, the shrinkage prior implies that higher dimensions have negligible variance and therefore are non-informative. The LSPCM eliminates the need for choosing a model selection criterion and only requires the fitting of a single model to simultaneously infer both the optimal number of clusters and the effective latent space dimension, reducing the required computation time. Additionally, the Bayesian framework naturally provides uncertainty quantification for the number of clusters and the number of effective dimensions.

The remainder of this article is structured as follows: Section \ref{sec:lspcm-model} describes the proposed LSPCM while Section \ref{sec:lspcm-inference} outlines the inferential process along with practical implementation details. Section \ref{sec:lspcm-sim} describes simulation studies conducted to explore the performance of the LSPCM in a variety of settings where the numbers of nodes, dimensions and clusters, as well as cluster volumes, vary. Section \ref{sec:lspcm-appl} applies the proposed LSPCM to two real Twitter network data sets: one concerning football players in the English premier league and one concerning the Irish political context. Section \ref{sec:lspcm-discussion} concludes and discusses potential extensions. R \citep{rcoreteam_2024_r} code with which all results presented herein were produced is freely available from the \href{https://gitlab.com/gwee95/lspm}{\texttt{lspm}} GitLab repository.

\section{The latent shrinkage position cluster model}
\label{sec:lspcm-model}
To cluster nodes in a network, we introduce the novel LSPCM which draws on and fuses together ideas from both the latent shrinkage position model (LSPM) of \cite{gwee_2022_a}, the LPCM  of \cite{handcock_2007_modelbased} and the sparse finite mixture models of \cite{malsinerwalli_2014_modelbased}.

\subsection{The latent shrinkage position model} \label{ssec:lspm}

Network data typically take the form of an $n \times n$ adjacency matrix, $\mathbf{Y}$, where $n$ is the number of nodes and entry $y_{i,j}$ denotes the relationship or edge between nodes $i$ and $j$. Self-loops are not permitted and thus the diagonal elements of $\mathbf{Y}$ are zero. Here we consider binary edges but a variety of edge types can be considered. 

Under the LPM \citep{hoff_2002_latent}, edges are assumed independent, conditional on the latent positions of the nodes. The sampling distribution is then
$$
    \mathbb{P}(\mathbf{Y}\mid\alpha, \mathbf{Z}) = \prod_{i \neq j} \mathbb{P}(y_{i,j}\mid\alpha, \mathbf{z}_i,\mathbf{z}_j)
$$
where $\mathbf{Z}$ is the
matrix of latent positions, with $\mathbf{z}_i$ denoting the latent position of node $i$; $\alpha$ is a global parameter that captures the overall connectivity level in the network. Denoting by $q_{i,j}$ the probability of an edge between nodes $i$ and $j$, i.e. $\mathbb{P}(y_{i,j} = 1\mid\alpha, \mathbf{z}_i,\mathbf{z}_j)$, a logistic regression model formulation is used where the log odds of an edge between nodes $i$ and $j$ depends on the Euclidean distance between their respective positions $\mathbf{z}_i$ and $\mathbf{z}_j$ in the latent space,
\begin{equation} \label{eq:lspcm-logitprob}
\log \frac{q_{i,j}}{1-q_{i,j}} = \alpha - \Vert \mathbf{z}_i-\mathbf{z}_j \Vert^2_2.  \end{equation} 
As in \cite{gollini_2016_joint} and \cite{dangelo_2019_latent}, in \eqref{eq:lspcm-logitprob} the distance is taken to be the squared Euclidean distance, giving higher weight to the latent space in the model than that induced by a linear Euclidean distance.

\cite{gwee_2022_a} propose the LSPM, a Bayesian nonparametric extension of the LPM that allows automatic inference on the number of the infinite latent space dimensions that are necessary to fully describe the network i.e., the number that are effective since they have non-negligible latent position variance. Under the LSPM, the latent positions are assumed to have a zero-centred Gaussian distribution with diagonal precision matrix $\mathbf{\Omega}$, whose entries $\omega_{\ell}$ denote the precision of the latent positions in dimension $\ell$, for $\ell=1, \ldots, \infty$. Spherical Gaussian distributions are assumed, as is standard in latent position models, since the likelihood is invariant to rotations of the latent space \citep{handcock_2007_modelbased}. The LSPM employs a multiplicative truncated gamma process (MTGP) prior on the precision parameters: the latent dimension $h$ has an associated shrinkage strength parameter $\delta_h$, where the cumulative product of $\delta_1$ to $\delta_{\ell}$ gives the precision $\omega_{\ell}$. An unconstrained gamma prior is assumed for $\delta_1$, while a truncated gamma distribution is assumed for the remaining dimensions to ensure shrinkage. Specifically, for $i = 1, \ldots, n$

\begin{equation}
\begin{aligned}  
    \mathbf{z}_{i} \sim \text{MVN}(\mathbf{0}, \mathbf{\Omega}^{-1}) 
\end{aligned}
\end{equation} 
\begin{equation}
\begin{aligned}  
    \mathbf{\Omega} = \begin{bmatrix} 
    \omega_{1}^{-1} & \dots & 0 \\
    \vdots & \ddots & \\
    0 &        & \omega_{\infty}^{-1}
    \end{bmatrix}  \qquad
    \omega_{\ell} = \prod_{h=1}^{\ell} \delta_{h} \mbox{ for } \ell = 1, \ldots, \infty  \\
\end{aligned}
\end{equation} 
\begin{equation}
\begin{aligned}  
\notag
    \delta_1 \sim \text{Gam}(a_1, b_1=1) \qquad
    \delta_{h} \sim \text{Gam}^{\text{T}}(a_2, b_2=1, t_2 = 1)  \mbox{ for }h  > 1. 
\end{aligned}
\end{equation}
Here $a_1$ and $b_1$ are the gamma prior's shape and rate parameters on the first dimension's shrinkage parameter, while $a_2$ is the shape parameter, $b_2$ is the rate parameter, and $t_2$ is the left truncation point (here set to 1) of the truncated gamma prior for dimensions $h > 1$. This MTGP prior results in an increasing precision and therefore a shrinking variance of the positions in the higher dimensions of the latent space. Further details regarding these hyperparameters and their choices are given in Appendix
\ref{app:hyperparam}. Under this MTGP prior, the LSPM is nonparametric with infinitely many dimensions, where unnecessary higher dimensions' variances are increasingly shrunk towards zero. Dimensions that have variance very close to zero will then have little meaningful information encoded in them as the distances between nodes will be close to zero. Thus, under the LSPM, the effective dimensions are those in which the variance is non-negligible \citep{gwee_2022_a}.

\subsection{The latent shrinkage position cluster model} \label{ssec:lspcm-model}

To infer the unknown number of clusters of nodes, we propose the novel latent shrinkage position cluster model (LSPCM) which results from fusing the LSPM \citep{gwee_2022_a}, the latent position cluster model (LPCM) of \cite{handcock_2007_modelbased} and the sparse finite mixture model framework of \cite{malsinerwalli_2014_modelbased, malsinerwalli_2017_identifying}. Like the LPCM, the proposed LSPCM assumes that latent positions arise from a finite mixture of $G$ spherical multivariate normal distributions; however, unlike the LPCM, the LSPCM assumes an infinite-dimensional latent space as in the LSPM. Further, the LSPCM employs a sparsity-inducing Dirichlet prior on the mixture weights, facilitating automatic inference on the number of clusters.

Specifically, the LSPCM assumes, for $i = 1, \ldots, n$,
\begin{gather} \label{eq:z_mvn}
\mathbf{z}_{i} \sim \sum_{g=1}^G \tau_g \text{MVN} (\bm{\mu}_g,\psi_g^{-1}\mathbf{\Omega}^{-1})    \\
\bm{\mu}_g |\mathbf{\Omega}  \sim \text{MVN}(\mathbf{0}, \xi \mathbf{\Omega}^{-1})  
\qquad \psi_g  \sim \text{Gam}(a_{\psi}, b_{\psi}) 
\\ 
\mathbf{\Omega} = \begin{bmatrix} 
    \omega_{1}^{-1} & \dots & 0 \\
    \vdots & \ddots & \\
    0 &        & \omega_{\infty}^{-1}
    \end{bmatrix} \qquad \omega_{\ell} = \prod_{h=1}^{\ell} \delta_{h} \mbox{ for }  \ell = 1, \ldots, \infty\\ 
    \delta_{1} \sim \text{Gam}(a_{1}, b_{1} = 1) \quad
    \delta_{h} \sim \text{Gam}^{\text{T}}(a_{2}, b_{2} = 1, t_2 = 1)  \mbox{ for }h  > 1. 
\end{gather}
Here $\tau_g$ denotes the probability that a node belongs to the $g$-th component, so that $\tau_g \geq 0$ $(g = 1,\ldots, G)$ and $\sum_{g=1}^G \tau_g =1$. 
For the mean latent position $\bm{\mu}_g$ of the $g$-th component, a multivariate Gaussian prior with zero mean and a scaled version of the same precision matrix as the latent positions with its MTGP prior is assumed; this ensures the latent positions will all shrink towards zero at higher dimensions. Consequently, the component means are increasingly shrunk towards zero, resulting in higher dimensions characterised by increasingly overlapping mixture components, and, as a result, becoming less informative for cluster separation. Further, as in \cite{ryan_2017_bayesian}, the prior covariance on $\bm{\mu}_g$ is inflated by a scaling factor $\xi$; values of $\xi>1$ result in cluster means that are more dispersed than cluster members. While a range of values of $\xi = \{1, 9, 100\}$ were considered, posterior predictive checks suggested that, similar to \cite{ryan_2017_bayesian},  $\xi = 9$ showed good model fit by encouraging both separate yet connected clusters. Additionally, a cluster specific factor $\psi_g$ scales the precision matrix of the latent positions providing the flexibility to model clusters of different volumes. 

While the LPCM assumes a finite mixture of spherical multivariate normal distributions and casts the problem of inferring the number of clusters as one of model selection, the LSPCM considers an overfitted (or sparse) finite mixture model which provides automatic inference on the number of clusters. Here, following \cite{malsinerwalli_2014_modelbased,fruhwirthschnatter_2019_handbook},  a symmetric Dirichlet prior is placed on the mixture weights $\bm{\tau}$ with a gamma prior placed on the Dirichlet's influential sparsity inducing hyperparameter $\nu$, i.e.,
$$\bm{\tau} = (\tau_1, \ldots, \tau_G) \sim \text{Dir}(\nu, \ldots, \nu) \qquad \nu \sim \text{Gam}(a_{\nu},G b_{\nu}).$$
Under such an overfitted mixture, a distinction is made between the number of mixture components $G$, the number of non-empty mixture components $G_+$, and the number of clusters in the data $G^*$.
A large number $G$ of initial components is specified and under a gamma prior that favours small values of $\nu$, unnecessary components are emptied out during the inferential process. The number of non-empty components $G_+$ then serves as an estimate of the number of clusters $G^*$. Thus, the LSPCM allows for automatic, simultaneous inference of the number of clusters and of the number of effective latent space dimensions, while requiring only a single model to be fit to the network data.

While the LSPM adopts a nonparametric approach, an overfitted sparse finite mixture approach for the clustering aspect of LSPCM is employed here as the sampling algorithm is simpler than that required to fit an infinite mixture model. Further, as detailed in \cite{frhwirthschnatter_2018_from}, under certain hyperparameter specifications, similar inference with respect to the number of clusters is obtained under the sparse finite and nonparametric approaches. Additionally, \cite{frhwirthschnatter_2018_from} and \cite{murphy_2020_infinite} highlight that overfitted
mixtures serve well in situations where the data arise from a moderate number of clusters, even with growing sample size, whereas infinite mixtures are suited to cases
where the number of clusters increases with sample size. As the networks considered here are anticipated to arise from a moderate number of clusters, an overfitted approach is preferred and taken.

\begin{figure}[htbp]
     \centering
     \begin{subfigure}[b]{0.57\linewidth}
         \centering
         \includegraphics[width=\linewidth]{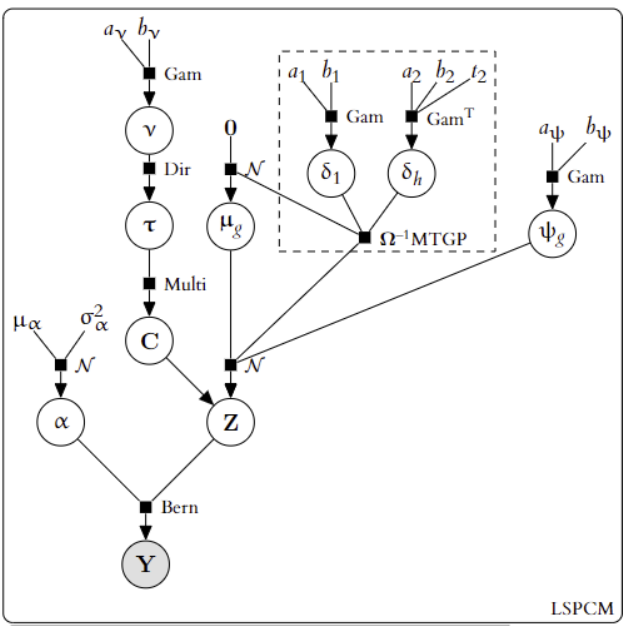}
         \caption{}
         \label{fig:lspcm_graph}
     \end{subfigure}
        \caption{The graphical model representation of the LSPCM.}
        \label{fig:graph}
\end{figure}

To facilitate inference, we introduce $\bm{C} = (\bm{c}_1, \ldots, \bm{c}_n)$ which contains the latent component membership vectors $\bm{c}_i = (c_{i1}, \ldots, c_{iG})$, for $i = 1, \ldots, n$, where $c_{ig} = 1$ if node $i$ belongs to component $g$ and $c_{ig} = 0$ otherwise. Thus, $\bm{c}_i \sim \text{MultiNom}(1, \bm{\tau})$. A summarising graphical model representation of the LSPCM is shown in Figure \ref{fig:graph}.

\section{Inference}
\label{sec:lspcm-inference}

Denoting by $\bm{\Theta}$ the component means $\bm{\mu}_1, \ldots, \bm{\mu}_G$, the precision scaling factors $\psi_1, \ldots, \psi_G$ and the precision matrix $\bm{\Omega}$,
the joint posterior distribution of the LSPCM is then
\begin{equation}
\begin{aligned}  
\notag
\mathbb{P}(\alpha, \mathbf{Z}, \mathbf{C}, \bm{\tau}, \bm{\Theta} \mid \mathbf{Y}) &\propto  \mathbb{P}(\mathbf{Y} \mid \alpha, \mathbf{Z}) \mathbb{P}(\alpha) \mathbb{P}(\mathbf{Z} \mid \bm{C}, \bm{\tau}, \bm{\Theta}) \mathbb{P}(\bm{C} \mid \bm{\tau}) \mathbb{P}(\bm{\tau}) \mathbb{P}(\bm{\Theta})
\end{aligned}
\end{equation}
where $\mathbb{P}(\alpha)$ is the non-informative $N(\mu_{\alpha} = 0,\sigma^2_{\alpha} = 4)$ prior for the $\alpha$ parameter, and $\mathbb{P}(\bm{\tau})$ and $\mathbb{P}(\bm{\Theta})$ denote the prior distributions outlined in Section \ref{ssec:lspcm-model}.
For the MTGP priors, similar to \cite{durante_2017_a} and \cite{gwee_2022_a}, we set the hyperparameters to $a_1 = 2 $ and $a_2 = 3$. A gamma hyperprior is used on the hyperparameter $\nu$ with shape $a_{\nu}=5$ and rate $b_{\nu}=5$ that is found to work well empirically in the simulation studies. For the gamma prior on $\psi_g$, as LPMs can only model networks of connected nodes, and as uncovering clustering structure is not possible if some clusters are so variable that they mask others, only values of $\psi_g$ near 1 are of interest. A strongly informed gamma prior centered on 1 is therefore assumed for $\psi_g$ with $a_\psi = 400$ and $b_\psi = 400$, giving a prior concentrated around 1. A summary of the LSPCM hyperparameter settings used is detailed in Appendix \ref{app:hyperparam}.

\subsection{An adaptive Metropolis-within-Gibbs sampler}
\label{sec:MwGsampler}

Markov chain Monte Carlo (MCMC) is employed to draw samples from the joint posterior distribution. After defining necessary notation in Appendix \ref{app:notation}, derivations of the full conditional distributions of the latent positions, cluster membership vectors and model parameters are given in Appendix \ref{app:fullconditionals}. As the MTGP prior is nonparametric and assumes an infinite number of latent dimensions, in practice setting an initial finite truncation level, $p_0$, on the number of dimensions fitted is required. Following \cite{bhattacharya_2011_sparse}, an adaptive Metropolis-within-Gibbs sampler is employed where the truncation dimension $p \leq p_0$ is dynamically shrunk or augmented as the sampler proceeds, with inferential validity ensured as detailed in \cite{bhattacharya_2011_sparse}.  The number of effective dimensions, $\hat{p}$, can then  be subsequently inferred (see Section \ref{sssec:postprocess}).

Denoting the current iteration by $s$ for $s = 1, \dots, S$ (but for clarity $s$ is omitted where it is unnecessary), a single pass of the sampler proceeds as follows: 

\begin{enumerate}
\item Sample the component mean $\bm{\mu}_g$ for $g = 1, \ldots, G$ from \\ $\text{MVN}_p\left(\frac{ \sum_{i=1}^n c_{ig}  \mathbf{z}_{i}}{ \sum_{i=1}^n c_{ig}  + \xi^{-1} } , \left[ \bm{\Omega} \left(\sum_{i=1}^n c_{ig}  + \xi^{-1} \right) \right]^{-1} \right).$ 

\item Sample $\check{\nu}$ from a Gam$(\sigma_{\nu}, \frac{\sigma_{\nu}}{\nu^{(s)}})$ where $\sigma_{\nu}$ is a step size factor. Accept $\check{\nu}$ as $\nu^{(s+1)}$ with probability $
\frac{p(\nu = \check{\nu}\mid -)}{p(\nu = \nu^{(s)} \mid -)} \frac{f(\check{\nu}; a_{\nu}, b_{\nu})}{f(\nu^{(s)}; a_{\nu}, b_{\nu})}$, where $p(\nu \mid -)$ is the full conditional of $\nu$ and $f(\cdot; a_{\nu}, b_{\nu})$ is the gamma density. Otherwise, set  $\nu^{(s+1)} =\nu^{(s)}$.
\item Sample the mixing weights $\bm{\tau}$ from Dir$(\sum_{i=1}^n c_{i1} + \nu, \ldots, \sum_{i=1}^n c_{iG} + \nu)$.
\item Sample the cluster-specific precision scaling parameter $\psi_g$ for $g = 1, \ldots, G$ from \\ $ \text{Gam} \left( a_{\psi}+\frac{p\sum_{i=1}^n c_{ig}}{2} \quad,  \quad b_{\psi}+ \frac{1}{2} \sum_{i=1}^n  (\mathbf{z}_{i} - \bm{\mu}_g)^T 
        \mathbf{\Omega} (\mathbf{z}_{i} - \bm{\mu}_g)c_{ig} \right)$.
\item Sample the latent memberships $\bm{c}_i$ for each node $i = 1, \ldots, n$ from a Multinomial with $G$ categories by drawing the $g$-th category with probability \\ $ \frac{\tau_g \phi_p(\mathbf{z}_i ; \bm{\mu}_g, \psi_g^{-1}\bm{\Omega}^{-1})}{\sum_{r=1}^G \tau_r \phi_p(\mathbf{z}_i ; \bm{\mu}_r, \psi_g^{-1}\bm{\Omega}^{-1})  }$ where $\phi_p(\mathbf{z}_i; \bm{\mu}, \psi_g^{-1}\bm{\Omega}^{-1})  $ is the $p$-dimensional multivariate normal density.
    \item Sample $\check{\mathbf{Z}}$ from a MVN$_p(\mathbf{Z}^{(s)}, k\psi_g^{-1}\mathbf{\Omega}^{-1(s)})$ proposal distribution where $k$ is a step size factor. Accept $\check{\mathbf{Z}}$ as $\mathbf{Z}^{(s+1)}$ with probability $\frac{\mathbb{P}(\mathbf{Y}\mid \alpha^{(s)}, \check{\mathbf{Z}})}{\mathbb{P}(\mathbf{Y}\mid\ \alpha^{(s)}, \mathbf{Z}^{(s)})} \frac{\phi_p(\mathbf{z}_i ; \check{\mathbf{Z}}, k\psi_g^{-1}\mathbf{\Omega}^{-1(s)})}{\phi_p(\mathbf{z}_i ;\mathbf{Z}^{(s)}, k\psi_g^{-1}\mathbf{\Omega}^{-1(s)})}$, otherwise set $\mathbf{Z}^{(s+1)} = \mathbf{Z}^{(s)}$. 
        
    \item Sample $\check{\alpha}$ from an informed Gaussian proposal distribution
    $N(\Bar{\mu}_{\alpha}, \Bar{\sigma}^2_{\alpha})$ \citep{gormley_2010_a} where
    \begin{equation}
\begin{aligned}  
\notag
\Bar{\mu}_{\alpha} = \alpha^{(s)} + \Bar{\sigma}^2_{\alpha}\times \left[ \sum_{i \neq j}y_{i,j} - \sum_{i \neq j} \hspace{2pt} \frac{\exp(\alpha^{(s)} - \Vert \mathbf{z}_i-\mathbf{z}_j \Vert^2_2)}{1+\exp(\alpha^{(s)} - \Vert \mathbf{z}_i-\mathbf{z}_j \Vert^2_2)}  + \frac{1}{\sigma^2_{\alpha}}(\mu_{\alpha}-\alpha^{(s)}) \right],
\end{aligned}
\end{equation}
and
\begin{equation}
\begin{aligned}  
\notag
\Bar{\sigma}^2_{\alpha} = \left\{\sum_{i \neq j} \hspace{2pt} \frac{\exp(\alpha^{(s)} - \Vert \mathbf{z}_i-\mathbf{z}_j \Vert^2_2)}{\left[1+\exp(\alpha^{(s)} - \Vert \mathbf{z}_i-\mathbf{z}_j \Vert^2_2)\right]^2} + \frac{1}{\sigma^2_{\alpha}}\right\}^{-1}.
\end{aligned}
\end{equation}
    Then, accept $\check{\alpha}$ as $\alpha^{(s+1)}$
    following the Metropolis-Hastings acceptance ratio.
    \item Sample $\delta_{1}$ from \\ 
    $\text{Gam} \left( \frac{ \left( n+G\right) p}{2} + a_1, \right. \frac{1}{2} \sum_{i=1}^{n} \sum_{g=1}^{G} (\mathbf{z}_{i} - \bm{\mu}_g)^T \psi_g^{-1}
    \left(\prod_{m=2}^{\ell} \delta_{m}\right)\mathbf{I}_p (\mathbf{z}_{i} - \bm{\mu}_g) c_{ig} + \\
    \left. \qquad \qquad \qquad \qquad \qquad \qquad \quad
    \frac{1}{2}\sum_{g=1}^{G}\bm{\mu}_g^T (\xi^{-1}) \left( \prod_{m=2}^{\ell} \delta_{m}\right)\mathbf{I}_p \bm{\mu}_g +b_1  \right).$ \\
        \item Sample $\delta_{h}^{(s+1)}$ for $h=2, \ldots, p$ from \\
$\text{Gam}^\text{T} \left( \frac{ \left( n+G\right)(p-h+1)}{2} + a_2, \right. \frac{1}{2} \sum_{i=1}^{n} (\mathbf{z}_{i} - \bm{\mu}_g)^T \psi_g^{-1}
    \left(\prod_{m=1, m \neq h}^{\ell} \delta_{m}^{(s^*)}\right)\mathbf{I}_p (\mathbf{z}_{i} - \bm{\mu}_g) c_{ig} + \\
    \left. \qquad \qquad \qquad \qquad \qquad \qquad \quad \frac{1}{2}\sum_{g=1}^{G}\bm{\mu}_g^T (\xi^{-1}) \left(\prod_{m=1, m \neq h}^{\ell} \delta_{m}^{(s^*)}\right)\mathbf{I}_p \bm{\mu}_g  + b_2, \quad 1 \right)$ \\
        where $s^* = s + 1$ for $m < h$ and $s^* = s$ for $m>h$. 
        
    \item Calculate $\omega_{\ell}$ by taking the cumulative product of $\delta_1$ to $\delta_{\ell}$.  

    \item At iteration $s$, adapt the truncation dimension $p^{(s)}$ with probability $\exp(-\kappa_0 - \kappa_1s)$ as detailed in Section \ref{sssec:adaptive}. 
\end{enumerate}

\subsection{Practical implementation details }
When implementing the adaptive Metropolis-within-Gibbs sampler, practical details such as initialisation of latent variables and parameters, adapting the truncation dimension and post-processing of the MCMC chain require attention.

\subsubsection{Initialisation } \label{sssec:initialisation}
While the LSPCM assumes an infinite number of latent dimensions and a potentially large number of components in the overfitted mixture, this is computationally impractical due to finite computational resources. Setting initial values for both the truncation level of the latent dimensions, $p_0$, and for the number of components in the mixture, $G$, prevents the sampler from exploring computationally unfeasible models. 
Here, $p_0 = 5$ and $G = 20$ are used throughout as these settings were empirically observed to strike a balance between exploring complex models and computational feasibility for the type of networks analysed.

The LSPCM's parameters, cluster allocations and latent positions also require initialisation, achieved here as follows, where $s = 0$:

\begin{enumerate}
    \item Calculate the geodesic distances \citep{kolaczyk_2020_statistical} between the nodes in the network. 
    Apply classical multidimensional scaling \citep{cox_1994_multidimensional} to the geodesic distances and set $\mathbf{Z}^{(0)}$ to be the resulting $n \times p_0$ positions. Empirically, this approach was more computationally efficient than setting $\mathbf{Z}^{(0)}$ to the LPM solution, with little notable loss in inferential performance.
    \item Fit a standard regression model, with regression coefficients $\alpha$ and $\beta$, to the vectorised adjacency matrix where $\mbox{log odds} (q_{i,j}=1) 
                  = \alpha - \beta \Vert \mathbf{z}_i^{(0)}-\mathbf{z}_j^{(0)}\Vert^2 _2 $ to obtain estimates $\hat{\alpha}$ and $\hat{\beta} $. Set $\alpha^{(0)} = \hat{\alpha}$. 
    \item As the LSPCM model \eqref{eq:lspcm-logitprob} constrains $\beta = 1$, centre and rescale the latent positions by setting $\mathbf{Z}^{(0)} = \sqrt{|\hat{\beta}|} \tilde{\mathbf{Z}}^{(0)}$, where $\tilde{\mathbf{Z}}^{(0)}$ are the mean-centred initial latent positions.
    \item Apply model-based clustering to $\mathbf{Z}^{(0)}$ via \texttt{mclust} \citep{scrucca_2016_mclust} with up to $G$ clusters and the \texttt{EEI} covariance structure to obtain initial cluster allocations.
    \item Obtain $\omega_{\ell}^{(0)}$ for $\ell=1, \ldots, p_0$ by calculating the empirical precision of each column of $\mathbf{Z}^{(0)}$.
    \item Set $\delta_1^{(0)} = \omega_{1}^{(0)}$ and calculate $\delta_h^{(0)} = \frac{\omega_{h}^{(0)}}{\omega_{h-1}^{(0)}}$ where $h = 2, \ldots, p_0$.
    \end{enumerate}

\subsubsection{Adapting the truncation dimension}\label{sssec:adaptive}
As in \cite{bhattacharya_2011_sparse} and \cite{gwee_2022_a}, the LSPCM adapts the truncation dimension $p$ as the MCMC sampler runs. This adaptation approach is employed to ease computational load as there are often relatively few important factors. Further, under the LSPCM, later dimensions will have very small, but not exactly zero, variance thus contributing to the distance between nodes in (\ref{eq:lspcm-logitprob}), and potentially introducing bias to parameter estimates. This behaviour is considered in \cite{gwee_2022_a}. Similarly to \cite{bhattacharya_2011_sparse}, the probability of adapting $p$ at iteration $s$ is taken as $\mathbb{P}(s) = \exp(-\kappa_0 - \kappa_1s)$, which decreases exponentially as the MCMC chain evolves. Here, setting $\kappa_0 = 4$ and $ \kappa_1 = 5 \times 10^{-4} $ demonstrated good performance in empirical experiments.  

After the burn-in period, at an adaptation step, when $p>1$, $p$ is reduced based on the cumulative proportion of the latent position variance that the dimensions $\ell = 1, \ldots , p - 1$ contain. If the dimensions up to the $\ell $-th cumulatively contain a proportion greater than $\epsilon _{1}$ of the total variance, then the dimensions from $\ell +1$ to $p$ add little information, and $p$ is reduced to $\ell $. In general, $\epsilon_1 = 0.8$ was found to work well but higher values of $\epsilon_1$ also showed good performance in networks with large $n$. If the criterion for reducing $p$ is not met, an increase in $p$ is then considered by examining $\delta_p^{-1}$ and a threshold $\epsilon_2$: if  $\delta_p^{-1} > \epsilon_2$, $p$ is increased by 1, with the additional associated parameters drawn from their respective priors. We consider $\epsilon_2 = 0.95$ which was found to work well in practice. The case where $p=1$ at an adaptation step requires a different criterion:  if the proportion of latent positions with absolute deviation from their empirical mean $> 1.96$ (the 95\% critical value of a standard Normal ) is greater $0.05 \times \epsilon_3$, then $p$ is increased to 2; setting $\epsilon_3 = 5$ was found to work well in practice. Reducing $p$ is not considered if $p = 1$. 

\subsubsection{Post-processing of the MCMC chain}\label{sssec:postprocess}

The likelihood function of the LSPCM depends on the Euclidean distances between the latent positions, hence it remains unaffected by rotations, reflections, or translations of these positions, giving rise to identifiability issues. To ensure valid posterior inference, similar to \cite{gormley_2010_a}, a Procrustean transformation of the sampled latent positions $\mathbf{Z}^{(1)}, \ldots, \mathbf{Z}^{(S)}$ is considered. The transformation aligns the sampled positions with a reference configuration $\tilde{\mathbf{Z}}$, which is selected based on the configuration that yields the highest log-likelihood during the burn-in phase of the MCMC chain. Although this choice is arbitrary, it has little effect as the reference configuration solely serves the purpose of addressing identifiability.

The MCMC sampler provides insight on the number of effective latent dimensions $\hat{p}$. Here, post hoc, the number of effective dimensions at each iteration $s$, $\hat{p}^{(s)}$, is determined as $\hat{p}^{(s)} = p^{(s)} - m^{(s)}$, where $m^{(s)}$ is the number of non-effective dimensions, i.e. dimensions with negligible variance.
For each iteration, the cumulative proportion of variance of each dimension is computed. If the first $\ell^{(s)}$ dimensions cumulatively contain a proportion greater than $\epsilon _{1} = 0.8$ of the total variance, then the dimensions from $\ell^{(s)} +1$ to $p^{(s)}$ are considered as non-effective and  $m^{(s)} = p^{(s)} - \ell^{(s)}$. 
Similar to \cite{bhattacharya_2011_sparse} $\hat{p}_m$, the mode of the thinned $\hat{p}^{(1)}, \ldots, \hat{p}^{(S)}$, is used to denote the modal effective dimension,  with empirical intervals quantifying the associated uncertainty.

Since the MCMC samples of the component allocations will have varying numbers of non-empty components, and because of the label switching problem caused by invariance of the mixture with respect to permutation of the component labels, care must be taken when deriving posterior summaries of cluster labels and parameters.
While \cite{fruhwirthschnatter_2019_handbook} give an overview of many potential approaches, here, as it demonstrated robust and accurate performance for the networks examined, we adopt the procedure of \cite{fritsch_2009_improved} to estimate the cluster labels of the nodes and subsequently the number of clusters. The method, implemented in the R package \texttt{mcclust} \citep{fritsch_2022_mcclust}, first estimates the posterior similarity matrix containing the proportion of times a pair of nodes are placed in the same cluster, and then maximises the posterior expected adjusted Rand index (PEAR) to obtain the optimal cluster labels.  In addition, as in \cite{malsinerwalli_2014_modelbased}, the posterior distribution of the number of filled components $G_+$ and the corresponding posterior mode $G_m$ are examined to inspect the uncertainty in the number of clusters. In some situations (see Section \ref{appl:football} for an example), the posterior mode may differ from the number of clusters estimated by the PEAR method. 
Finally, for summarizing the posterior distributions of the cluster parameters, as in \cite{gormley_2010_a}, cluster parameters are obtained after permuting cluster labels to minimise a loss function based on the cluster means.

\section{Simulation studies}
\label{sec:lspcm-sim}
The performance of the LSPCM is assessed on simulated data scenarios by evaluating its ability to recover the  effective latent dimension of the latent space and to correctly infer the latent positions, the number of clusters and the cluster allocations. The latent positions are simulated according to \eqref{eq:z_mvn} and the mixture weights $\bm{\tau}$ are generated from a symmetric $\text{Dir}(10, \ldots, 10)$ which gives rise to clusters with similar numbers of nodes. Given the latent positions, a network is then generated as in \eqref{eq:lspcm-logitprob}. Three scenarios are considered: Sections \ref{ssec:n50p2g3} and \ref{ssec:n200p4g7}, 
respectively, assess the performance of LSPCM on networks with small and moderate numbers of nodes, dimensions, and clusters, while Section \ref{ssec:n50p2g3psi} assesses the performance when clusters have varying volumes i.e., when $\psi_g \neq \psi_g'$ for $g \neq g'$. A total of 30 networks are simulated in each scenario. In Sections \ref{ssec:n50p2g3} and \ref{ssec:n50p2g3psi}, the MCMC chains are run for 500,000 iterations with a burn-in of 50,000 iterations, thinned every 1,000th iteration while in Section \ref{ssec:n200p4g7}, 1,000,000 iterations are considered, with a burn-in of 100,000 iterations, thinned every 2,000th iteration due to the larger settings considered. 
To ensure acceptance rates in the $20\%-40\%$ range, step sizes of $\sigma_{\nu} = 0.5$ and $k$ values between 1 and 2.5 were employed in the proposal distributions.

\subsection{Scenario 1: small networks} \label{ssec:n50p2g3}

Networks with $n=50$ are generated with the true number of latent dimensions $p^*=2$ and true number of clusters $G^*=3$, with shrinkage strength $\bm{\delta}=(1, 1.05)$, $\psi_1=\psi_2=\psi_3=1$, and cluster mean positions $\bm{\mu} =\{(0,0), (-4,0), (-4,4)\}$ giving well separated clusters. 
Across the 30 simulated networks, the smallest cluster contained 6 nodes while the largest cluster contained 31, and fixing $\alpha = 6$ gave rise to network densities between 27\% and 38\%.

\begin{figure}[htbp]
     \centering
     \begin{subfigure}[b]{0.495\linewidth}
         \centering
         \includegraphics[width=\linewidth]{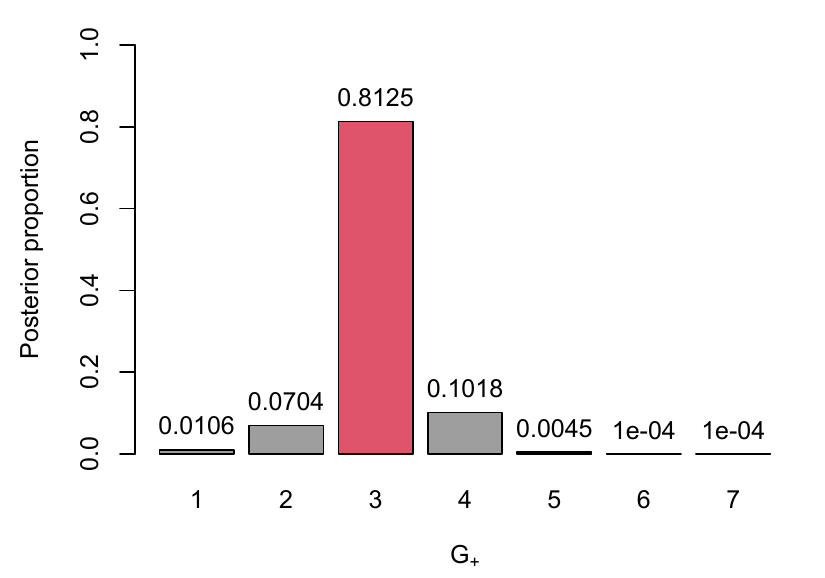}
         \caption{}
         \label{fig:result_n50_d2_cluster_bar}
     \end{subfigure}
     \hfill
     \begin{subfigure}[b]{0.495\linewidth}
         \centering
         \includegraphics[width=\linewidth]{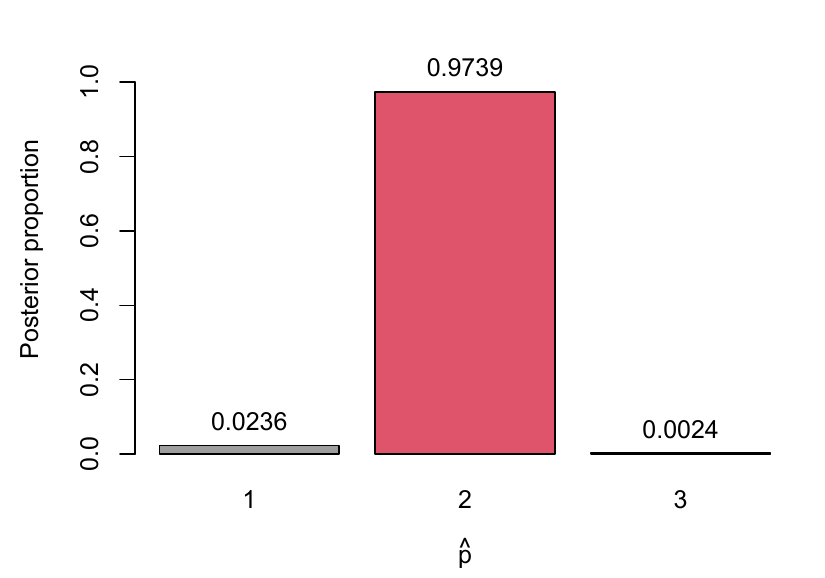}
         \caption{}
         \label{fig:result_n50_d2_dim_bar}
     \end{subfigure}
        \caption{(a) Posterior distribution of the number of active clusters $G_+$ and (b) the distribution of the effective latent space dimension $\hat{p}$ across 30 small simulated networks. The true number of clusters and latent dimensions are highlighted in red.}
        \label{fig:result_n50_d2}
\end{figure}

Figure \ref{fig:result_n50_d2_cluster_bar} shows that the posterior modal number of clusters $G_m = 3$ and Figure \ref{fig:result_n50_d2_dim_bar} shows that the modal effective number of dimensions is $\hat{p}_m=2$. Overall, the adjusted Rand index \citep[ARI,][]{hubert_1985_comparing}, which measures agreement between the inferred cluster memberships and the true cluster labels, shows good correspondence, with mean value of 0.88 (standard deviation (sd) = 0.11). Procrustes correlations \citep[PC,][]{peresneto_2001_how} between inferred and true latent positions also show that the latent positions are accurately estimated, with average value of $ 0.97$ (sd = 0.03). Additional visualisation of the posterior distributions on the variances, shrinkage strengths, and the parameter $\alpha$ are available in Appendix \ref{app:simstudies}.

Additionally, in this scenario sensitivity to the setting of $\epsilon_1$ was explored by also considering $\epsilon_1 = 0.9$ when deciding to add a truncation dimension but with the lower value of 0.8 employed when inferring the effective number of dimensions; there was no change in the modal effective dimension inferred, with only a small change in the posterior probability of $\hat{p} = 3$ to 0.9225.

For comparison, the LPCM is fitted on the same simulated data using the R package \texttt{latentnet} \citep{krivitsky_2008_fitting}. Upon fitting 25 LPCMs with different combinations of $p = \{1, \ldots, 5\}$ and $G = \{1, \ldots, 5\}$ to each of the 30 simulated networks, the BIC suggested 2 dimensions and 3 clusters for all of them, with 
average ARI and PC values between the  LPCM's posterior modal cluster labels and the true labels  of $0.87 \,(\mbox{sd} =0.12)$  
and $0.96 \,(\mbox{sd} = 0.04)$, respectively. In terms of computational cost, fitting the LSPCM on a computer with an i9-13900H CPU and 32GB RAM took on average 18.5 minutes. The LPCM with the correct $p$ and $G$ took 5.55 minutes on average to run for the same number of iterations. Across the various simulated networks, the total time taken to fit the 25 LPCMs with different combinations of $p$ and $G$ was comparable to or longer than the time taken to fit the LSPCM. However, notably, when fitting the LPCMs no quantification of the uncertainty in $p$ and $G$ is provided.

\subsection{Scenario 2: moderately sized networks} \label{ssec:n200p4g7}
Networks with $n=200$ were generated with $p^*=3$ latent dimensions and $\bm{\delta}=(1,1.1,1.05)$. The true number of clusters $G^*=7$ with $\bm{\mu} =\{(-5,0,0), (-5,5,0),(0,-5,5),(0,0,-5),$ $ (2,0,2), (-2,2,-2),(0,-2,0)\}$ and $\psi_g= 1$ for $g = 1, \ldots, 7$. The 7 clusters have different degrees of separation across the dimensions. Among the 30 simulated networks, the smallest cluster had 4 nodes while the largest had 54. The parameter $\alpha=20$, which resulted in networks with density varying between 23\% and 31\%.

\begin{figure}[htbp]
     \centering
     \begin{subfigure}[b]{0.495\linewidth}
         \centering
         \includegraphics[width=\linewidth]{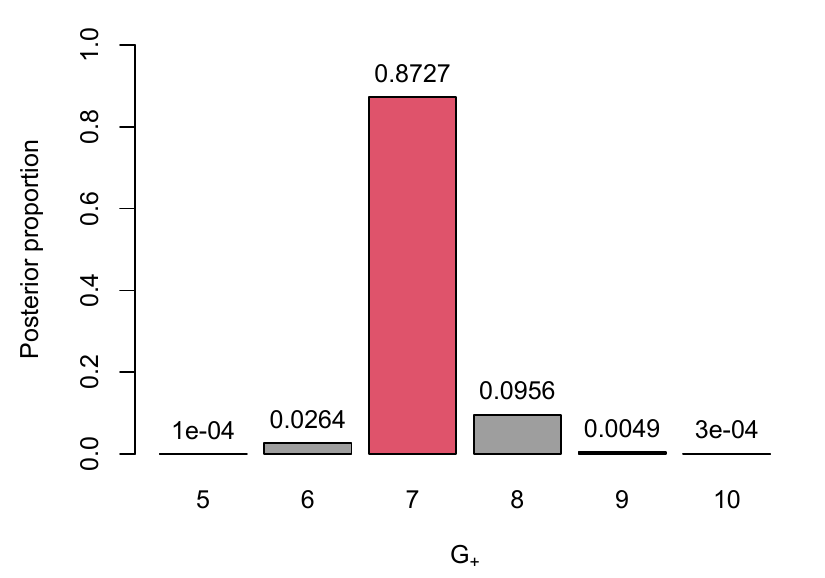}
         \caption{}
         \label{fig:result_n200_d4_cluster_bar}
     \end{subfigure}
     \hfill
     \begin{subfigure}[b]{0.495\linewidth}
         \centering
         \includegraphics[width=\linewidth]{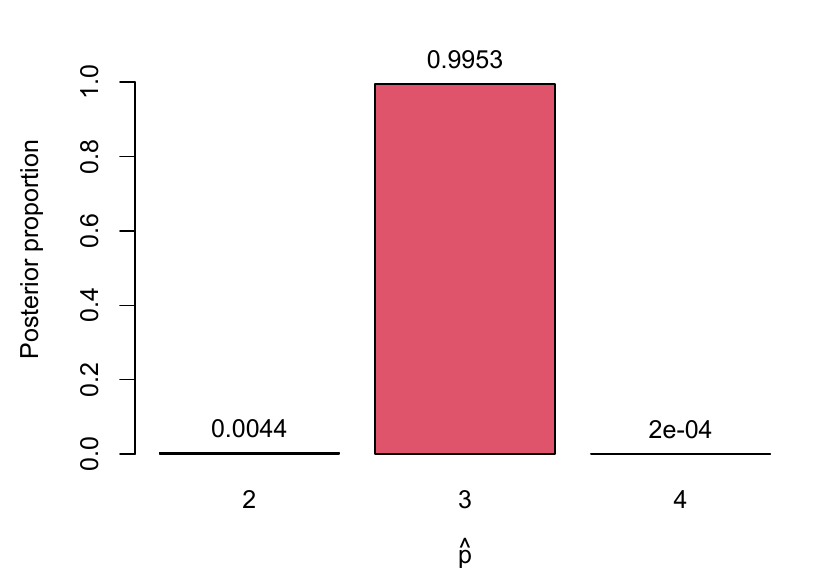}
         \caption{}
         \label{fig:result_n200_d4_dim_bar}
     \end{subfigure}
        \caption{(a) Posterior distribution of the number of active clusters $G_+$ and (b) the distribution of the effective latent space dimension $\hat{p}$ across 30 moderately sized simulated networks. The true number of clusters and latent dimensions are highlighted in red.}
        \label{fig:result_n200_d4}
\end{figure}

Figure \ref{fig:result_n200_d4_cluster_bar} shows that the posterior modal number of clusters is correctly inferred with $G_m = 7$ and Figure \ref{fig:result_n200_d4_dim_bar} suggests the modal effective number of dimensions is also correctly inferred with $\hat{p}_m=3$.

The ARI values indicate very accurate and robust inference of cluster membership with mean value of 0.941 (sd=0.024). The average PC value was 0.994 (sd=0.002). The PC values are calculated using only dimensions up to and including the modal effective number of dimensions; in cases where $\hat{p}_m < p^*$, the higher dimensions are not included in the comparison resulting in lower PC values. Additional visualisation of the posterior distributions of the variances, shrinkage strengths, and the parameter $\alpha$  for this setting are available in Appendix \ref{app:simstudies}. In scenarios with larger $p^*$, underestimation of the true number of dimensions often occurred as empirically with only $n=200$ and $G^*=7$ the relative variance in later dimensions was intuitively small.

Upon fitting the LPCM with 3 dimensions and 7 clusters the average ARI and PC values between the LPCM's posterior modal cluster labels and the true labels were $0.84 \,(0.09)$  and  $0.98 \,(0.01)$, respectively (standard deviation in brackets). The average time for the completion of one MCMC chain for the LSPCM was 149 minutes. However, fitting the LPCM for multiple combinations of $p$ and $G$ would incur a considerably larger computational cost. In fact, solely for $p = 3$ and $G = 7$ it took 142 minutes to run the MCMC via \texttt{latentnet} for the same number of iterations, and again no uncertainty quantification is provided.

\subsection{Scenario 3: networks with clusters of different volumes } 
\label{ssec:n50p2g3psi}

Networks are generated with the same settings as in Section \ref{ssec:n50p2g3} except with clusters of different volumes. Two settings are considered: clusters of slightly different volumes where $\psi_g$ values are set as $(\psi_1, \psi_2, \psi_3) = (\frac{4}{5}, 1, \frac{5}{4})$ and clusters of highly different volumes with $(\psi_1, \psi_2, \psi_3) = (\frac{1}{5}, 1, 5)$. For comparison purposes, the networks simulated in Section \ref{ssec:n50p2g3} will be considered as networks with clusters of the same volume as $\psi_g = 1 \forall g$. For the setting with slightly different volumes, network densities varied between 26\% and 38\% while in the highly different volumes setting, networks had densities between 22\% and 38\%. 

\begin{figure}[htbp]
     \centering
     \begin{subfigure}[b]{0.495\linewidth}
         \centering
         \includegraphics[width=\linewidth]{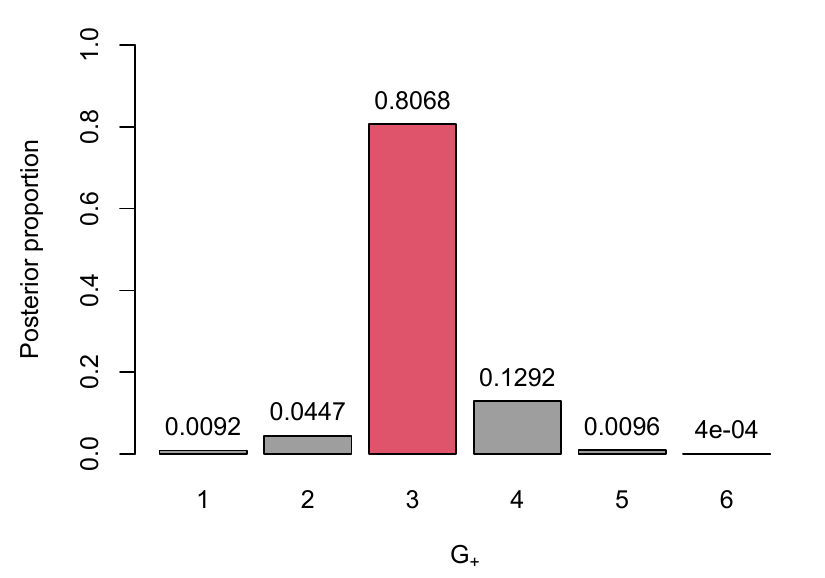}
         \caption{}
         \label{fig:result_n50_d2_psi_small_cluster_bar}
     \end{subfigure}
     \hfill
     \begin{subfigure}[b]{0.495\linewidth}
         \centering
         \includegraphics[width=\linewidth]{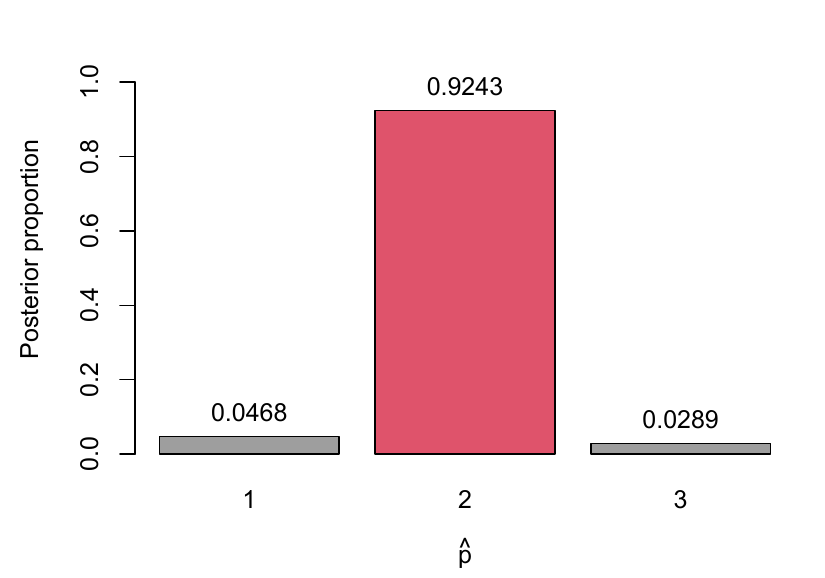}
         \caption{}
         \label{fig:result_n50_d2_psi_small_dim_bar}
     \end{subfigure}
        \caption{(a) Posterior distribution of the number of active clusters $G_+$ and (b) the distribution of the effective latent space dimension $\hat{p}$ across 30 small simulated networks with clusters of slightly different volumes. The true number of clusters and dimensions are highlighted in red.}
        \label{fig:result_n50_d2_psi_small}
\end{figure}

\begin{figure}[htbp]
     \centering
     \begin{subfigure}[b]{0.495\linewidth}
         \centering
         \includegraphics[width=\linewidth]{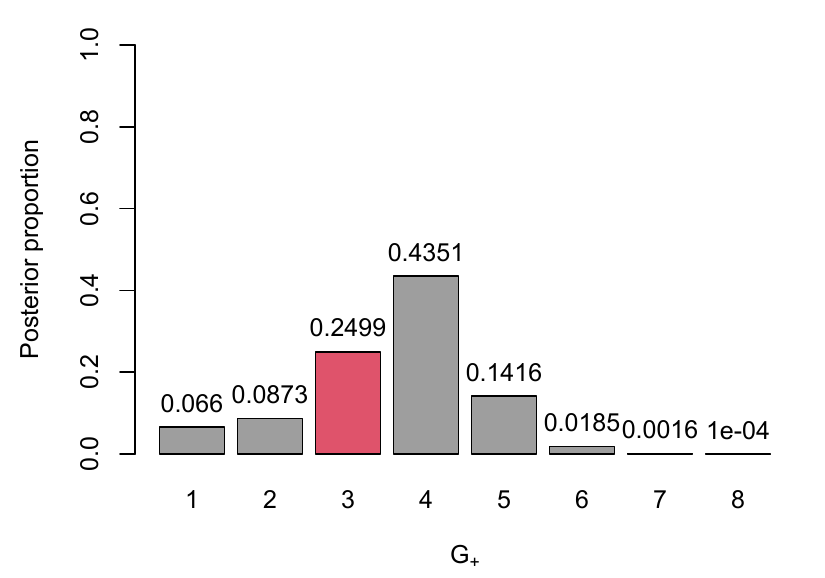}
         \caption{}
         \label{fig:result_n50_d2_psi_big_cluster_bar}
     \end{subfigure}
     \hfill
     \begin{subfigure}[b]{0.495\linewidth}
         \centering
         \includegraphics[width=\linewidth]{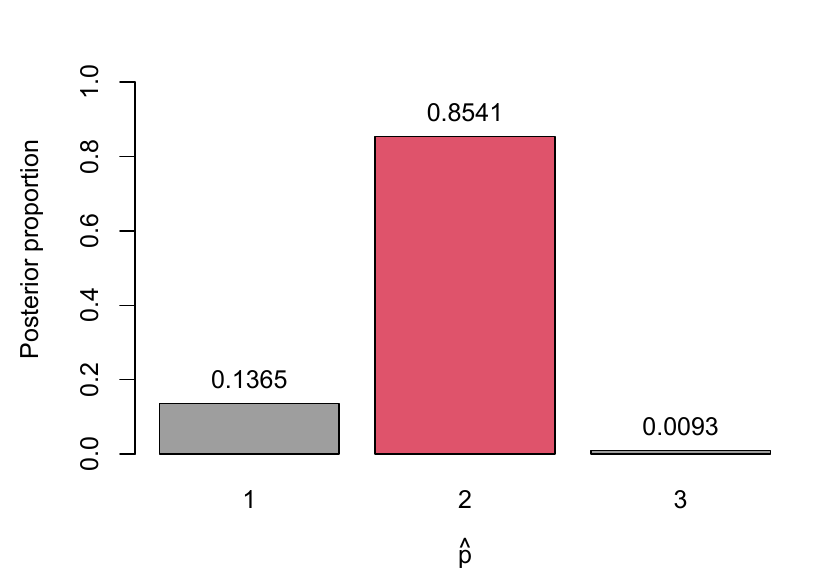}
         \caption{}
         \label{fig:result_n50_d2_psi_big_dim_bar}
     \end{subfigure}
        \caption{(a) Posterior distribution of the number of active clusters $G_+$ and (b) the distribution of the effective latent space dimension $\hat{p}$ across 30 small simulated networks with clusters of highly different volumes. The true number of clusters and dimensions are highlighted in red.}
        \label{fig:result_n50_d2_psi_big}
\end{figure}

Figure \ref{fig:result_n50_d2_psi_small} shows that, for clusters of slightly different volumes, the number of clusters and effective dimensions were correctly inferred, but with lower certainty than shown in Figure \ref{fig:result_n50_d2} for networks with clusters of the same volume. For the highly different volumes setting, Figure \ref{fig:result_n50_d2_psi_big_cluster_bar} shows the posterior modal number of clusters $G_m=4$ which underestimates the true number of clusters, while \ref{fig:result_n50_d2_psi_big_dim_bar} shows the effective number of dimensions is correctly inferred but again with less certainty than in the setting with clusters of the same volume. For networks with clusters of highly different volumes, intuitively the ARI values indicate poor clustering performance with mean value of 0.617 (sd=0.208); for the slightly different volumes setting the average ARI was 0.882 (sd=0.105). Similarly, the average PC values were 0.899 (sd=0.091) and 0.973 (sd=0.032) for the highly and slightly different volumes settings, respectively.

\begin{figure}[htbp]
     \centering
     \begin{subfigure}[b]{0.495\linewidth}
         \centering
         \includegraphics[width=\linewidth]{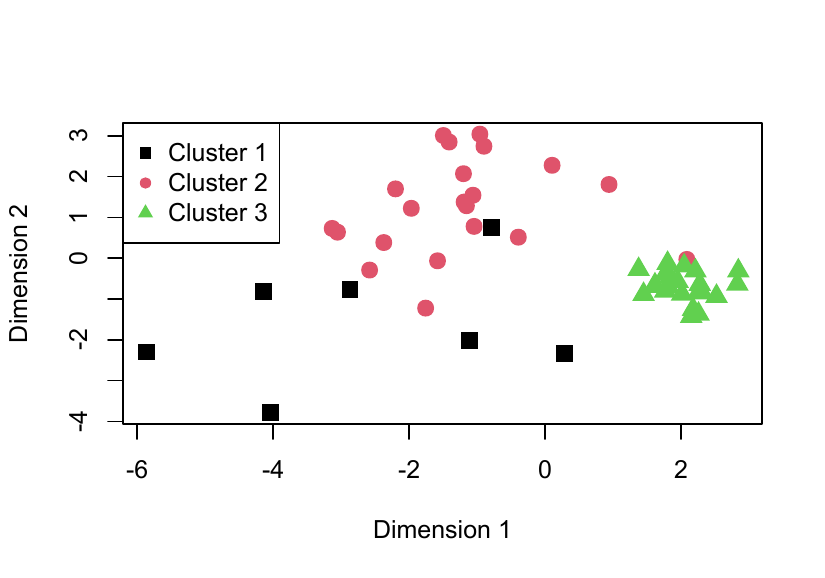}
         \caption{}
         \label{fig:result_psi_pos_true}
     \end{subfigure}
     \hfill
     \begin{subfigure}[b]{0.495\linewidth}
         \centering
         \includegraphics[width=\linewidth]{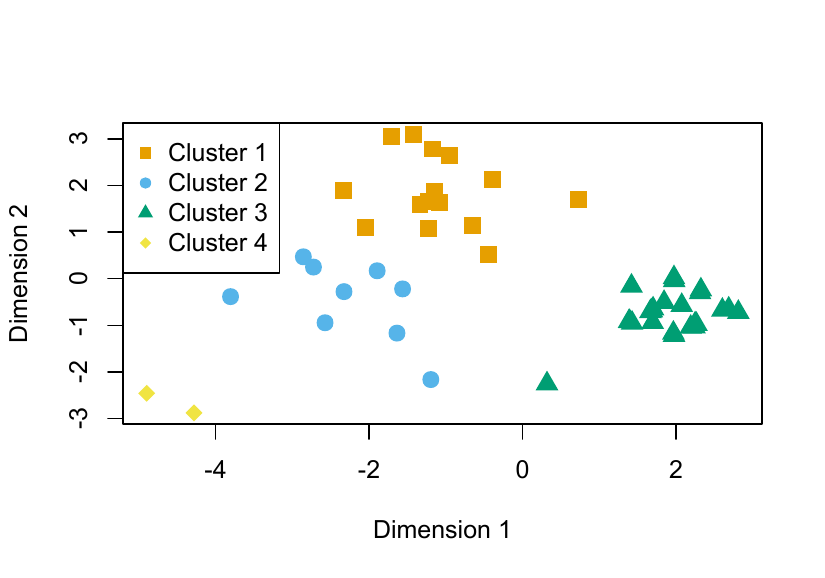}
         \caption{}
         \label{fig:result_psi_pos_lspcm}
     \end{subfigure}     
        \caption{One of the 30 simulated networks when clusters have highly different volumes with (a) the true positions and cluster labels and (b) the LSPCM posterior mean positions conditioned on the modal effective number of dimensions ($\hat{p}_{m}=2$) and the estimated cluster labels.}
        \label{fig:result_psi_pos}
\end{figure}

Figure \ref{fig:result_psi_pos_true} highlights the intuitive difficulty in clustering nodes in networks with clusters of highly different volumes. The nodes from the cluster with the largest volume (indicated by black squares in Figure \ref{fig:result_psi_pos_true}) are far apart in the latent space, and many lie closer to nodes from another cluster e.g., to nodes from the medium volume cluster indicated by red circles. The result, as illustrated in Figure \ref{fig:result_psi_pos_lspcm}, is that the LPSCM intuitively fits more clusters with smaller volumes than the truth.

\section{Application to Twitter network data}
\label{sec:lspcm-appl}

The LSPCM is fitted to two binary Twitter networks with different characteristics: a football players network with a small number of nodes, in which each player is known to belong to one of three football clubs, and a network among Irish politicians with a moderate number of nodes, where each politician is affiliated to one of seven Irish political parties. The data are publicly available at this \href{http://mlg.ucd.ie/aggregation/}{\texttt{link}} \citep{greene_2013_producing}. In the analyses, hyperparameters, initial values, and step sizes are set as in Section \ref{sec:lspcm-sim}. 

\subsection{Football players network} \label{appl:football}
This binary network consists of directed edges indicating the presence of Twitter mentions from one English Premier League football player's Twitter account to another. The data were adapted from those provided in \cite{greene_2013_producing} by considering only the top 3 clubs that have the most player Twitter accounts: 55 players playing for 3 different Premier League clubs are considered with 15 players from Stoke City football club (Stoke), 23 players from Tottenham Hotspur (Spurs), and 17 players from West Bromwich Albion (West Brom). In total, there were 497 mentions between players, giving a network density of 16.73\%. 
To fit the LSPCM to these data, 10 MCMC chains are run, each for 1,000,000 iterations with a burn-in period of 100,000 and thinned every 1,000th. The average time for the completion of one MCMC chain was 30.6 minutes. With 10 MCMC chains, the Gelman-Rubin convergence criterion \citep{gelman_2014_bayesian}, computed conditional on the  modal effective number of dimensions, was satisfied with $1.0 \leq \hat{R} < 1.1$ for the $\alpha$ , $\nu $, and $\psi$ parameters, and with $\hat{R}=1.3 $ and $\hat{R}=1.2$ for $\delta_1$ and $\delta_2$ respectively. Trace plots are provided in Appendix \ref{app:application}.

Figure \ref{fig:footballbar} illustrates that under the LSPCM the posterior modal number of clusters  $G_m =$ 1 (1, 4) and the modal effective number of dimensions is $\hat{p}_m=2$ (1, 2), with 95\% empirical intervals in brackets. There is uncertainty in $G_+$ as demonstrated by the wide posterior interval and notable support for $G_+ = 3$, but there is strong certainty on the number of effective dimensions $\hat{p}$.

\begin{figure}[htbp]
     \centering
     \begin{subfigure}[b]{0.49\linewidth}
         \centering
         \includegraphics[width=\linewidth]{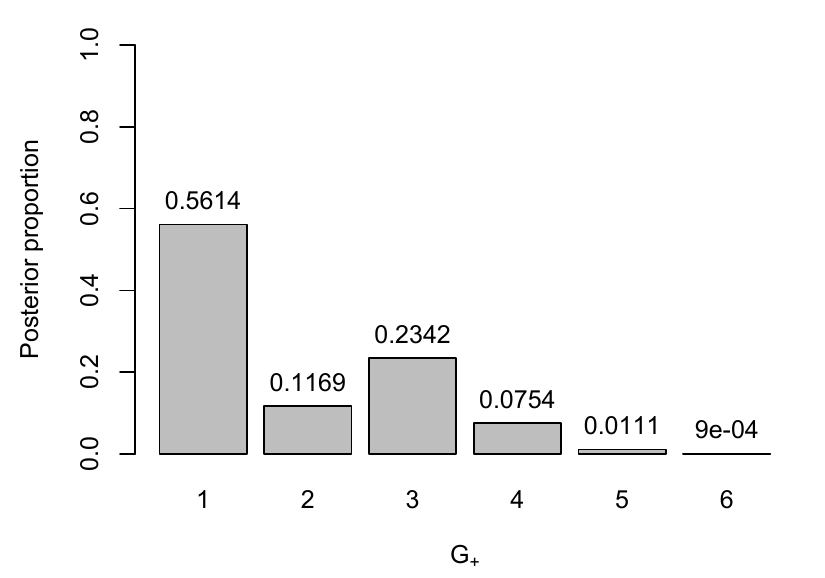}
         \caption{}
         \label{fig:football_cluster_bar}
     \end{subfigure}
     \hfill
     \begin{subfigure}[b]{0.49\linewidth}
         \centering
         \includegraphics[width=\linewidth]{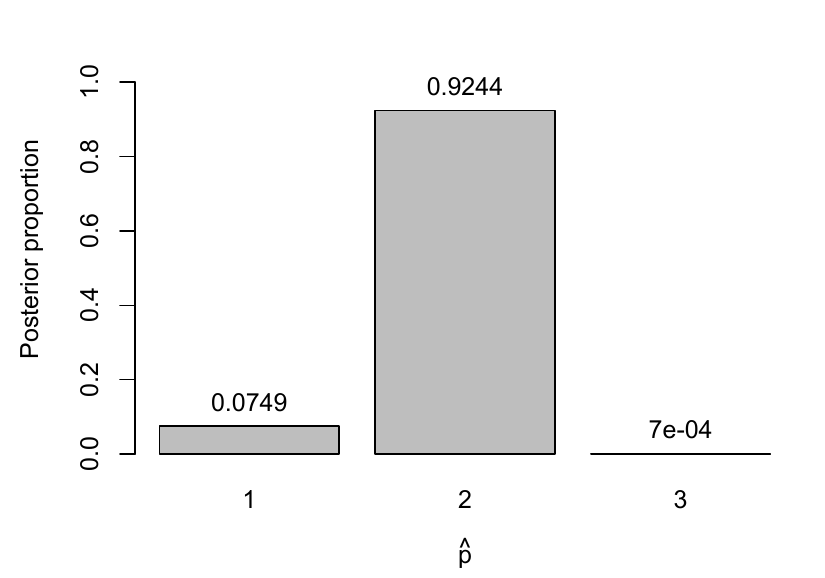}
         \caption{}
         \label{fig:football_dim_bar}
     \end{subfigure}
        \caption{For the football players network, (a) the posterior distribution of the number of non-empty components $G_+$, and (b) the distribution of the number of effective latent space dimensions $\hat{p}$.}
        \label{fig:footballbar}
\end{figure}

\begin{figure}[htbp]
     \centering
     \begin{subfigure}[b]{0.31\linewidth}
         \centering
         \includegraphics[width=\linewidth]{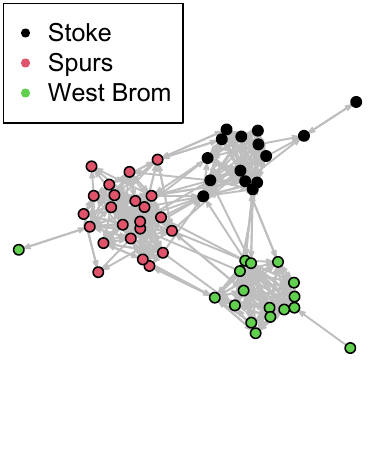}
         \caption{}
         \label{fig:football_true}
     \end{subfigure}
     \hfill
     \begin{subfigure}[b]{0.31\linewidth}
         \centering
         \includegraphics[width=\linewidth]{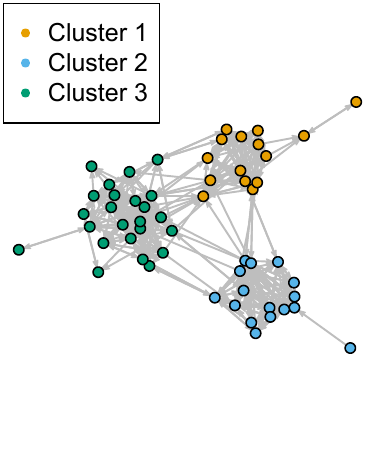}
         \caption{}
         \label{fig:football_lspcm}
     \end{subfigure}
    \begin{subfigure}[b]{0.33\linewidth}
         \centering
         \includegraphics[width=\linewidth]{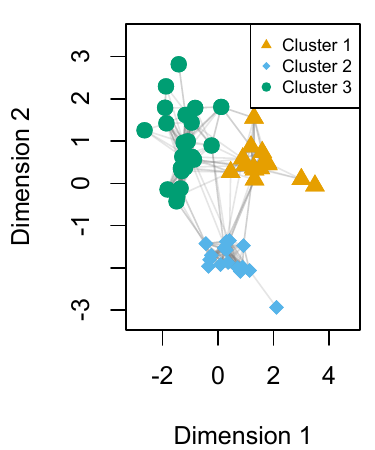}
         \caption{}
         \label{fig:football_lspcm_pos}
     \end{subfigure}
        \caption{Football players network, (a) the Fruchterman-Reingold layout with players coloured by club, (b) the Fruchterman-Reingold layout with players coloured by inferred cluster label and (c) posterior mean latent positions on the $\hat{p}_m = 2$ effective latent dimensions coloured by inferred cluster label.}
        \label{fig:footballpos}
\end{figure}

Despite $G_m =1$, 
maximising PEAR across the 10 chains resulted in an estimate of 3 clusters, with an ARI of 0.94 between this final clustering and the players' clubs. Figures \ref{fig:football_true} and \ref{fig:football_lspcm} show the Fruchterman-Reingold layouts \citep{kolaczyk_2020_statistical} of the network with the players' clubs and inferred clusters detailed respectively. Only one player was clustered differently to the other players in his club: the West Brom player Romelu Lukaku who was originally from Chelsea football club and was on a season-long loan deal with West Brom. Lukaku mentioned and was mentioned by only one player, the Spurs player Jan Vertonghen and intuitively the LSPCM has clustered them together. 
Figure \ref{fig:football_lspcm_pos} shows the posterior mean latent positions coloured by the cluster labels that maximises the PEAR across the 10 chains. Cluster 1 (which contains all of the Stoke players only) and cluster 2 (all players from Spurs, and Romelu Lukaku) are separate from each other on the first dimension, while in the second dimension cluster 3 (which captures West Brom players except Romelu Lukaku) is separated from clusters 1 and 2, indicating both effective latent dimensions are necessary for representation of the nodes in the three clusters.

\begin{figure}[htbp]
   \centering
   \includegraphics[width=7.5cm, height=6cm]{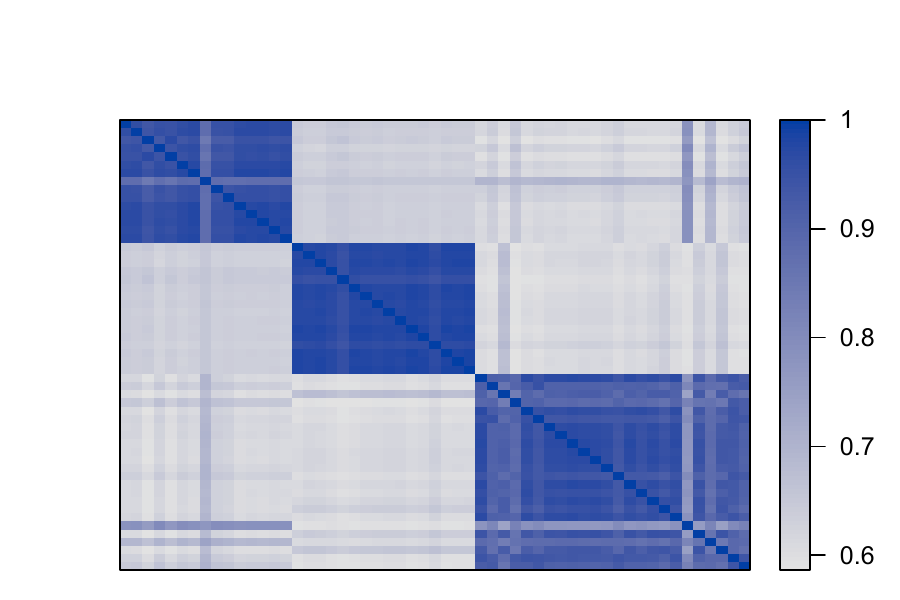}
   \caption{Heat map of the posterior similarity matrix of the cluster labels inferred from the football network, ordered by the cluster labels.}
   \label{fig:football_heat}
\end{figure}

Through the posterior similarity matrix, Figure \ref{fig:football_heat} illustrates the uncertainty in the football players' cluster labels where the colour intensity indicates the probability of a player being clustered together with another player. Cluster 2 (predominantly players from West Brom) has the strongest certainty in its cluster labels, while there is some uncertainty in the membership of players in clusters 1 and 3.

Upon fitting 16 LPCMs from all combinations of $p = \{1, \ldots, 4\}$ and $G = \{1, \ldots, 4\}$, the BIC suggests 3 clusters and 2 dimensions as optimal. The ARI between the cluster labels of the optimal LPCM and LSPCM is 1, indicating that both approaches provide the same cluster allocations.
The average Procrustes correlation between the posterior mean latent positions under the optimal LPCM and the LSPCM was 0.973 (standard deviation of 0.005) across the 10 LSPCM chains. Additional results regarding the posterior distributions of variance and shrinkage strength parameters can be found in the Appendix \ref{app:application}.


\subsection{Irish politicians network} \label{app:irish}

The LSPCM is used to analyse a Twitter network between 348 Irish politicians from  2012. The network consists of binary directed edges indicating if one politician follows another \citep{greene_2013_producing}. Each of the politicians is affiliated with one of the seven Irish political parties: 49 are affiliated with Fianna Fáil (FF), 143 with Fine Gael (FG), 7 with the Green Party (Green), 79 with the Labour Party (Lab), 31 with Sinn Féin (SF), 8 with the United Left Alliance (ULA) and 31 are Independent (Ind). There are 16,856 directed relationships between the politicians, giving a network density of 13.96.
For inference, 10 MCMC chains are run, each for 2,000,000 iterations with a burn-in period of 100,000 and thinned every 4,500th sample. Here, the setting of $G=20$ and $b_{\nu}=5$ resulted in very many sparsely populated clusters which provided little clustering insight. Motivated by the context, $G=10$ and $b_{\nu}=1000$ were instead used resulting in fewer, more populated clusters the results of which are presented here; results under $G=20$ and $b_{\nu}=5$, and $G=10$ and $b_{\nu}=100$ are provided in Appendix \ref{app:application} for the interested reader. The average time for the completion of one MCMC chain was 11.71 hours. With 10 MCMC chains, the Gelman-Rubin convergence criterion \citep{gelman_2014_bayesian}, computed conditional on the modal effective number of dimensions, was satisfied with $1.0 \leq \hat{R} < 1.1$ for the $\alpha $, $\nu $, and $\psi_g$ parameters, and $1.0 \leq \hat{R} < 1.6$ for $\bm{\delta}$ parameters. Trace plots are provided in Appendix \ref{app:application}.

Figure \ref{fig:ie_bar} shows that, under the LSCPM, the posterior modal number of clusters is 6 (5, 7) and the modal number of effective dimensions is 4 (3, 4), with the 95\% empirical intervals reported in brackets. Upon fitting multiple LPCMs across all combinations of $p = \{2, \ldots, 8\}$ and $G = \{2, \ldots, 8\}$, the BIC suggests 6 clusters and  6 dimensions. The ARI between the cluster labels of the optimal LPCM and LSPCM is 0.55, indicating different clustering solutions under the 2 approaches. The mean Procrustes correlation between the posterior mean positions under the LPCM and LSPCM was 0.71 (standard deviation of 0.01) across the 10 LSPCM chains. Additional results regarding the posterior distributions of variance and shrinkage strength parameters can be found in Appendix \ref{app:application}.

\begin{figure}[htbp]
     \centering
     \begin{subfigure}[b]{0.49\linewidth}
         \centering
         \includegraphics[width=\linewidth]{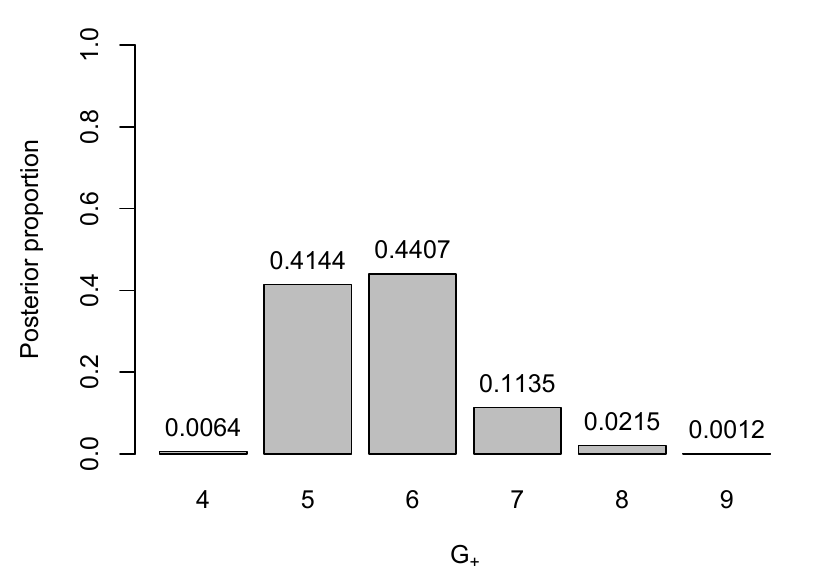}
         \caption{}
         \label{fig:ie_cluster_bar}
     \end{subfigure}
     \hfill
     \begin{subfigure}[b]{0.49\linewidth}
         \centering
         \includegraphics[width=\linewidth]{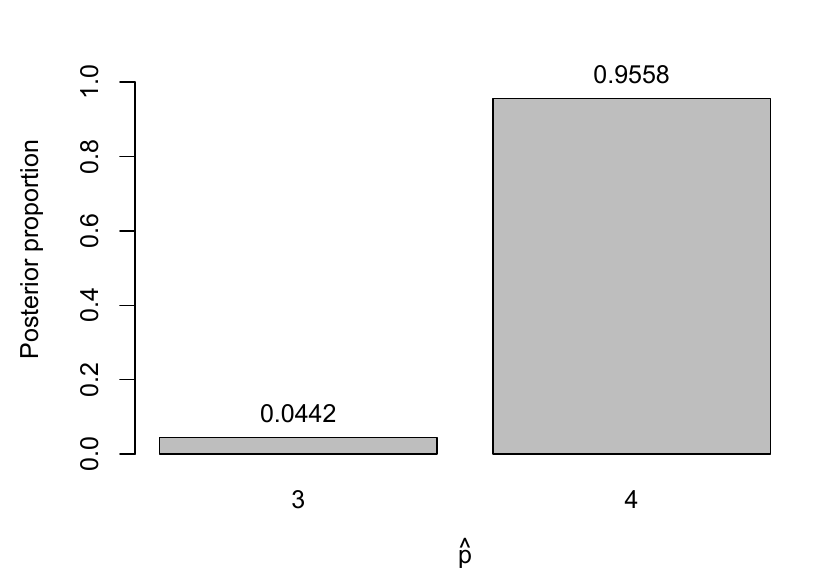}
         \caption{}
         \label{fig:ie_dim_bar}
     \end{subfigure}
        \caption{For the Irish politicians network: (a) the posterior distribution of the number of non-empty components $G_+$, and (b) the distribution of the effective latent space dimension $\hat{p}$.}
        \label{fig:ie_bar}
\end{figure}

\begin{figure}[tb]
     \centering
     \begin{subfigure}[b]{0.45\linewidth}
         \centering
         \includegraphics[width=\linewidth]{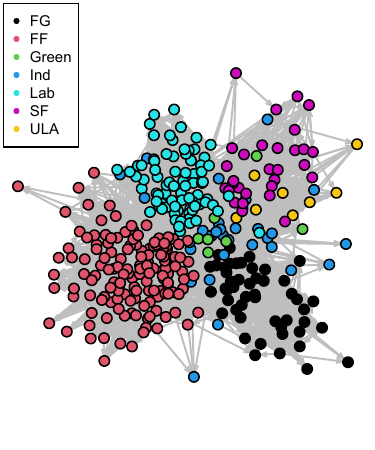}
         \caption{}
         \label{fig:ie_true}
     \end{subfigure}
     \hfill
     \begin{subfigure}[b]{0.45\linewidth}
         \centering
         \includegraphics[width=\linewidth]{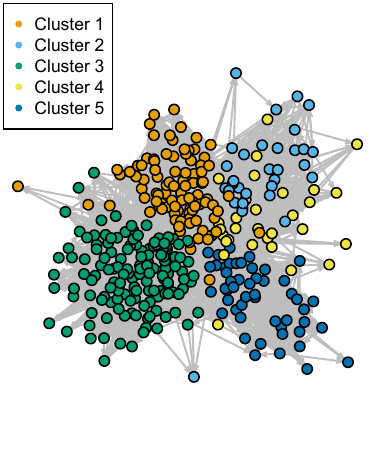}
         \caption{}
         \label{fig:ie_lspcm}
         \hfill
    \end{subfigure}
        \caption{Fruchterman-Reingold layout of the Irish politicians network with nodes coloured by (a) political party affiliation and (b) LSPCM inferred  cluster labels.}
        \label{fig:ie_pos}
\end{figure}

\begin{figure}[htbp]
   \centering
   \includegraphics[width=10cm, height=8cm]{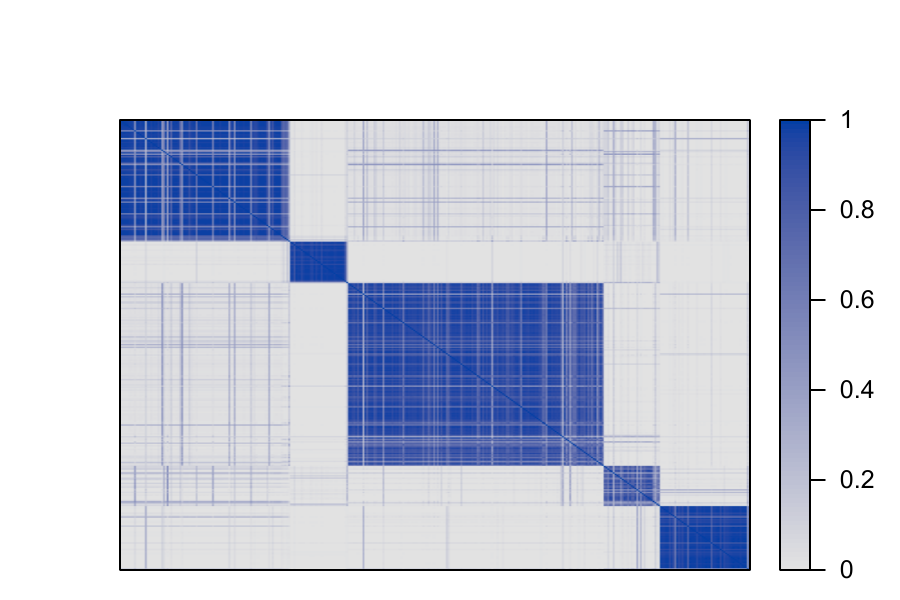}
   \caption{Heat map of the posterior similarity matrix of the Irish politicians ordered by the cluster labels.}
   \label{fig:ie_heat}
\end{figure}

Despite $G_m =6$, 
maximising the PEAR across the 10 chains, resulted in 5 clusters and the ARI between the inferred cluster labels and the politicians' political affiliations is 0.89.
Table \ref{tab:ie_cross} presents the cross-tabulation of the political party membership and the LSPCM cluster labels. Additionally, Figure \ref{fig:ie_pos} shows the Fruchterman-Reingold layout of the network with nodes coloured by political party affiliation and LSPCM inferred  cluster label.
The inferred clustering structure captures the composition and nature of the 2012 Irish political landscape: a coalition government of Fine Gael and the Labour Party were in power with Fianna Fáil the main opposition party. Intuitively, cluster 3 captures the large number of Fine Gael politicians with cluster 1 capturing the Labour politicians in power with them at the time. Cluster 5 captures all of the opposition's Fianna Fáil politicians. Cluster 2 solely contains the Sinn Féin candidates, with cluster 4 capturing many independent candidates and those from the United Left Alliance. Through the posterior similarity matrix, Figure \ref{fig:ie_heat} shows that politicians in cluster 2 (Sinn Féin) have the strongest certainty in their cluster labels while other politicians have some uncertainty.

\begin{table}[hbt!]
\centering
\caption{Cross-tabulation of political party membership and the LSPCM representative cluster labels. }
\label{tab:ie_cross}
\begin{tabular}{llccccc}
\toprule
 && \multicolumn{5}{c}{\textbf{Cluster}}\\
& & \em 1  & \em 2  & \em 3   & \em 4  & \em 5 \\ \midrule
                   & \em Fine Gael            & 3  &  & 140 &  &  \\
                   & \em Fianna Fáil          &   &   &    &  & 49 \\
\textbf{Political} & \em Labour Party         & 79 &   &    &   &  \\
\textbf{Parties}   & \em Independent          & 9  & 1 & 1 & 20  &  \\
                   & \em Green Party          & 3  &   &    & 3  & 1 \\
                   & \em Sinn Féin            &   & 31 &    &   &  \\
                   & \em United Left Alliance &   &   &    & 8  &  \\
\bottomrule
\end{tabular}
\end{table}

\begin{figure}[htbp] 
     \centering
    \begin{subfigure}[t]{0.99\linewidth}
     \includegraphics[width=\linewidth]{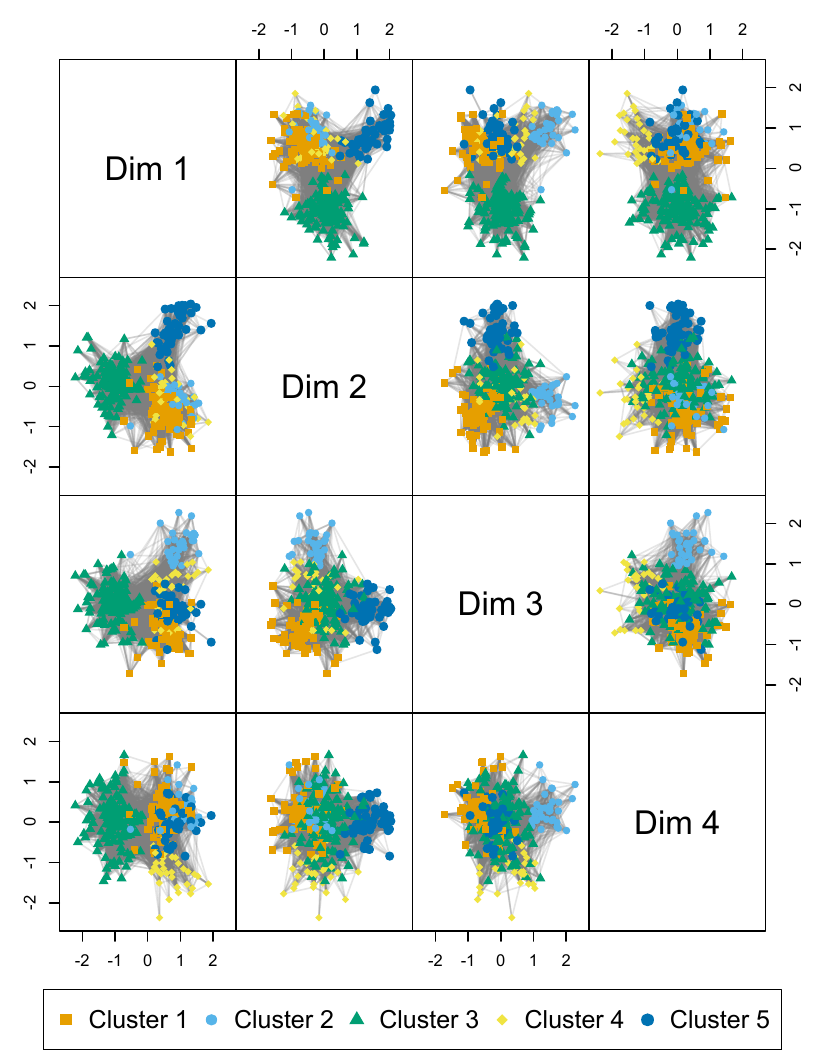}
     \end{subfigure}
     \caption{LSPCM inferred posterior mean latent positions of the Irish politicians on the $\hat{p}_m = 4$ dimensions with nodes coloured by cluster membership.}
     \label{fig:ie_pairs}
\end{figure}

Figure \ref{fig:ie_pairs} shows the politicians' posterior mean latent positions on the $\hat{p}_m = 4$ dimensions with nodes coloured by the maximised PEAR estimated cluster allocation across the 10 chains. On each dimension, one cluster tends to be located separately to the others, indicating that all the dimensions provide relevant information to distinguish the clusters. For example, on dimension 1 the politicians in cluster 3 are located separately from the other politicians. Similar patterns are apparent for politicians in cluster 5 on dimension 2, politicians in cluster 2 on dimension 3 and politicians in cluster 4 on dimension 4. Visually, the combination of all four dimensions allows for separation of the five clusters.

\section{Discussion}
\label{sec:lspcm-discussion}

The latent shrinkage position cluster model (LSPCM) enables simultaneous inference on the number of clusters of nodes in a network and the effective dimension of the latent space required to represent it. This is achieved, respectively, through employing a sparse prior on the weights of a mixture model and a Bayesian nonparametric shrinkage prior on the variance of the nodes' positions in the latent space. The LSPCM eliminates the need for the computationally expensive procedure of fitting multiple latent position cluster models with different numbers of clusters and dimensions, and then choosing the best model using a range of model selection criteria. The performance of the LSPCM was assessed through simulation studies and its application to sporting and political social networks uncovered interesting and intuitive clustering structures.

While the LSPCM demonstrated good performance, it is important to consider its sensitivity to specification of the parameters and threshold parameters employed. As the focus here is on the clustering solution, it is particularly important to consider sensitivity to the parameters of the gamma prior on the Dirichlet parameter, especially in small networks, but the context of the network at hand can provide some guidance.
Additionally, sensitivity to the latent position dimension threshold settings should be considered carefully by practitioners as underestimating the dimension could result in invalid inference while overestimating the effective number of dimensions could result in bias. In practice, the values required to recover the effective dimension may change according to the size and density of the network. Using a prior may enable more robust inference on this parameter. Regarding identifiability, as in the case of LPCM, the interplay between the number of dimensions and the number of clusters is of interest. Here, due to the shrinkage of the variance of the latent dimensions induced by the prior, higher dimensions are less likely to exhibit clustering structure as the cluster means become closer in higher dimensions. Thus, any clustering structure should be captured in the earlier dimensions.  Moreover, while the influence of the parameters of the Dirichlet prior and the MTGP shrinkage prior are intuitively related as they are specified within the same model, there is perhaps surprisingly little interaction between the dimension of the latent space and the number of clusters, as observed in \cite{handcock_2007_modelbased}. Relatedly, assessing model fit in a Bayesian clustering setting for network data is challenging; efforts to do so via posterior predictive checks \citep{
gwee_2022_a} and cross validation \citep{Sosa_Rodríguez_2021} have received attention, albeit in a non-clustering setting.

Fitting the LSPCM is computationally feasible on networks of the scale considered here, however, it would be computationally burdensome to apply it to larger networks. As the number of nodes increases, their computational cost increases quadratically \citep{saltertownshend_2013_variational, rastelli_2018_computationally}, making the LSPCM challenging to scale especially with the use of Metropolis-within-Gibbs sampling. \cite{spencer_2022_faster} propose faster inference for the LPM by combining Hamiltonian Monte Carlo and firefly Monte Carlo, which could be adapted for the LSPCM. Further, while adapting the LSPCM to facilitate modelling of networks with more complex edge types is a natural extension, such advances would come with 
additional computational cost. Addressing these issues could be possible by employing case-control approaches for the likelihood function \citep{raftery_2012_fast} and/or avoiding MCMC through the use of variational inference methods \citep{saltertownshend_2013_variational, gwee_2024}. Many variational algorithms have shown empirical success, and recent theoretical contributions by \cite{yang_2020_alpha} and \cite{liu_2023_variational} have provided better understanding into the statistical guarantees of these approaches.

Finally, while here a MTGP shrinkage prior was employed on the latent dimensions' variances and an overfitted mixture was used to infer the number of clusters, the LSPCM could be viewed as a member of a broader family of such models given the variety of alternative shrinkage priors and clustering approaches available. \cite{grushanina_2023_a} provides a broad review of approaches to infinite factorisations. For example, the Indian buffet process (IBP) has been employed to penalise increasing dimensionality in latent factor models; while it allows for the contribution from a dimension to be exactly zero \citep{knowles_2011_nonparametric, rokov_2016_fast}, the sparsity the IBP enforces could be too restrictive here. The spike-and-slab approach of the cumulative shrinkage prior of \cite{legramanti_2020_bayesian} is an alternative, likely fruitful approach. From the clustering point of view, an infinite mixture model could be used to infer the number of clusters. \cite{frhwirthschnatter_2018_from} discuss linkages between infinite and overfitted mixtures, where they highlight overfitted and infinite mixtures yield comparable clustering performance when the hyperpriors are matched.

\section*{Acknowledgement}
The authors are grateful for discussions with members of the Working Group in Model-based Clustering and for the reviewers' suggestions which greatly contributed to this work.

\paragraph{Funding Statement}
This publication has emanated from research conducted with the financial support of Research Ireland under Grant number 18/CRT/6049. For the purpose of Open Access, the author has applied a CC BY public copyright licence to any Author Accepted Manuscript version arising from this submission. 

\paragraph{Competing Interests}


None


\bibliographystyle{apalike}
\bibliography{arXiv/example.bib}{} 

\begin{thebibliography}{}

\bibitem[Aliverti and Durante, 2019]{aliverti_2019_spatial}
Aliverti, E. and Durante, D. (2019).
\newblock Spatial modeling of brain connectivity data via latent distance models with nodes clustering.
\newblock {\em Statistical Analysis and Data Mining: The ASA Data Science Journal}, 12:185--196.

\bibitem[Bhattacharya and Dunson, 2011]{bhattacharya_2011_sparse}
Bhattacharya, A. and Dunson, D.~B. (2011).
\newblock Sparse {B}ayesian infinite factor models.
\newblock {\em Biometrika}, 98:291--306.

\bibitem[Cox and Cox, 1994]{cox_1994_multidimensional}
Cox, T.~F. and Cox, M. A.~A. (1994).
\newblock {\em Multidimensional scaling}.
\newblock Chapman \& Hall.

\bibitem[Durante, 2017]{durante_2017_a}
Durante, D. (2017).
\newblock A note on the multiplicative gamma process.
\newblock {\em Statistics \& Probability Letters}, 122:198--204.

\bibitem[Durante and Dunson, 2014]{durante_2014_nonparametric}
Durante, D. and Dunson, D.~B. (2014).
\newblock Nonparametric {B}ayes dynamic modelling of relational data.
\newblock {\em Biometrika}, 101:883--898.

\bibitem[D’Angelo et~al., 2023]{dangelo_2023_modelbased}
D’Angelo, S., Alf\`o, M., and Fop, M. (2023).
\newblock Model-based clustering for multidimensional social networks.
\newblock {\em Journal of the Royal Statistical Society Series A: Statistics in Society}, 186:481--507.

\bibitem[D’Angelo et~al., 2019]{dangelo_2019_latent}
D’Angelo, S., Murphy, T.~B., and Alfò, M. (2019).
\newblock Latent space modelling of multidimensional networks with application to the exchange of votes in {E}urovision song contest.
\newblock {\em The Annals of Applied Statistics}, 13.

\bibitem[Erd\H{o}s and R\'enyi, 1959]{erds_1959_on}
Erd\H{o}s, P. and R\'enyi, A. (1959).
\newblock {\em On random graphs. I}, volume~6.
\newblock Publicationes Mathematicae.

\bibitem[Fritsch, 2022]{fritsch_2022_mcclust}
Fritsch, A. (2022).
\newblock mcclust: Process an mcmc sample of clusterings.

\bibitem[Fritsch and Ickstadt, 2009]{fritsch_2009_improved}
Fritsch, A. and Ickstadt, K. (2009).
\newblock Improved criteria for clustering based on the posterior similarity matrix.
\newblock {\em Bayesian Analysis}, 4:367--391.

\bibitem[Fruhwirth-Schnatter et~al., 2019]{fruhwirthschnatter_2019_handbook}
Fruhwirth-Schnatter, S., Celeux, G., and Robert, C.~P. (2019).
\newblock {\em Handbook of mixture analysis}.
\newblock CRC Press.

\bibitem[Frühwirth-Schnatter and Malsiner-Walli, 2018]{frhwirthschnatter_2018_from}
Frühwirth-Schnatter, S. and Malsiner-Walli, G. (2018).
\newblock From here to infinity: Sparse finite versus {D}irichlet process mixtures in model-based clustering.
\newblock {\em Advances in Data Analysis and Classification}, 13:33--64.

\bibitem[Gelman et~al., 2014]{gelman_2014_bayesian}
Gelman, A., Carlin, J.~B., Stern, H.~S., Dunson, D.~B., Vehtari, A., and Rubin, D.~B. (2014).
\newblock {\em Bayesian data analysis}.
\newblock Chapman \& Hall/Crc.

\bibitem[Gilbert, 1959]{gilbert_1959_random}
Gilbert, E.~N. (1959).
\newblock Random graphs.
\newblock {\em The Annals of Mathematical Statistics}, 30:1141--1144.

\bibitem[Gollini and Murphy, 2016]{gollini_2016_joint}
Gollini, I. and Murphy, T.~B. (2016).
\newblock Joint modeling of multiple network views.
\newblock {\em Journal of Computational and Graphical Statistics}, 25:246--265.

\bibitem[Gormley and Murphy, 2010]{gormley_2010_a}
Gormley, I.~C. and Murphy, T.~B. (2010).
\newblock A mixture of experts latent position cluster model for social network data.
\newblock {\em Statistical Methodology}, 7:385--405.

\bibitem[Greene and Cunningham, 2013]{greene_2013_producing}
Greene, D. and Cunningham, P. (2013).
\newblock Producing a unified graph representation from multiple social network views.
\newblock In {\em Proceedings of the 5th Annual ACM Web Science Conference}, WebSci '13, page 118–121, New York, NY, USA. Association for Computing Machinery.

\bibitem[Grushanina, 2023]{grushanina_2023_a}
Grushanina, M. (2023).
\newblock A review of {B}ayesian methods for infinite factorisations.

\bibitem[Gwee et~al., 2024a]{gwee_2022_a}
Gwee, X.~Y., Gormley, I.~C., and Fop, M. (2024a).
\newblock A latent shrinkage position model for binary and count network data.
\newblock {\em Bayesian Analysis}, page To appear.

\bibitem[Gwee et~al., 2024b]{gwee_2024}
Gwee, X.~Y., Gormley, I.~C., and Fop, M. (2024b).
\newblock Variational inference for the latent shrinkage position model.
\newblock {\em Stat}, pages 1 -- 22.

\bibitem[Handcock et~al., 2007]{handcock_2007_modelbased}
Handcock, M.~S., Raftery, A.~E., and Tantrum, J.~M. (2007).
\newblock Model-based clustering for social networks.
\newblock {\em Journal of the Royal Statistical Society: Series A (Statistics in Society)}, 170:301--354.

\bibitem[Hoff et~al., 2002]{hoff_2002_latent}
Hoff, P.~D., Raftery, A.~E., and Handcock, M.~S. (2002).
\newblock Latent space approaches to social network analysis.
\newblock {\em Journal of the American Statistical Association}, 97:1090--1098.

\bibitem[Holland et~al., 1983]{holland_1983_stochastic}
Holland, P.~W., Laskey, K.~B., and Leinhardt, S. (1983).
\newblock Stochastic blockmodels: First steps.
\newblock {\em Social Networks}, 5:109--137.

\bibitem[Hubert and Arabie, 1985]{hubert_1985_comparing}
Hubert, L. and Arabie, P. (1985).
\newblock Comparing partitions.
\newblock {\em Journal of Classification}, 2:193–218.

\bibitem[Jo et~al., 2021]{jo_2021_a}
Jo, W., Chang, D., You, M., and Ghim, G.-H. (2021).
\newblock A social network analysis of the spread of {COVID}-19 in {S}outh {K}orea and policy implications.
\newblock {\em Scientific Reports}, 11.

\bibitem[Kaur et~al., 2023]{kaur_2023_latent}
Kaur, H., Rastelli, R., Friel, N., and Raftery, A.~E. (2023).
\newblock Latent position network models.

\bibitem[Knowles and Ghahramani, 2011]{knowles_2011_nonparametric}
Knowles, D. and Ghahramani, Z. (2011).
\newblock Nonparametric {B}ayesian sparse factor models with application to gene expression modeling.
\newblock {\em The Annals of Applied Statistics}, 5.

\bibitem[Kolaczyk and Cs\'ardi, 2020]{kolaczyk_2020_statistical}
Kolaczyk, E.~D. and Cs\'ardi, G. (2020).
\newblock {\em Statistical analysis of network data with R}.
\newblock Springer, 2nd ed edition.

\bibitem[Krivitsky and Handcock, 2008]{krivitsky_2008_fitting}
Krivitsky, P.~N. and Handcock, M.~S. (2008).
\newblock Fitting position latent cluster models for social networks withlatentnet.
\newblock {\em Journal of Statistical Software}, 24.

\bibitem[Krivitsky et~al., 2009]{krivitsky_2009_representing}
Krivitsky, P.~N., Handcock, M.~S., Raftery, A.~E., and Hoff, P.~D. (2009).
\newblock Representing degree distributions, clustering, and homophily in social networks with latent cluster random effects models.
\newblock {\em Social Networks}, 31:204--213.

\bibitem[Legramanti et~al., 2020]{legramanti_2020_bayesian}
Legramanti, S., Durante, D., and Dunson, D.~B. (2020).
\newblock Bayesian cumulative shrinkage for infinite factorizations.
\newblock {\em Biometrika}, 107:745--752.

\bibitem[Liu and Chen, 2023]{liu_2023_variational}
Liu, Y. and Chen, Y. (2023).
\newblock Variational inference for latent space models for dynamic networks.
\newblock {\em Statistica Sinica}, 32.

\bibitem[Malsiner-Walli et~al., 2014]{malsinerwalli_2014_modelbased}
Malsiner-Walli, G., Frühwirth-Schnatter, S., and Grün, B. (2014).
\newblock Model-based clustering based on sparse finite {G}aussian mixtures.
\newblock {\em Statistics and Computing}, 26:303--324.

\bibitem[Malsiner-Walli et~al., 2017]{malsinerwalli_2017_identifying}
Malsiner-Walli, G., Frühwirth-Schnatter, S., and Grün, B. (2017).
\newblock Identifying mixtures of mixtures using {B}ayesian estimation.
\newblock {\em Journal of Computational and Graphical Statistics}, 26:285--295.

\bibitem[Murphy et~al., 2020]{murphy_2020_infinite}
Murphy, K., Viroli, C., and Gormley, I.~C. (2020).
\newblock Infinite mixtures of infinite factor analysers.
\newblock {\em Bayesian Analysis}, 15.

\bibitem[Ng et~al., 2021]{ng_2021_modeling}
Ng, J., Murphy, T.~E., Westling, T., McCormick, T.~H., and Fosdick, B.~K. (2021).
\newblock Modeling the social media relationships of {I}rish politicians using a generalized latent space stochastic blockmodel.
\newblock {\em The Annals of Applied Statistics}, 15.

\bibitem[Passino and Heard, 2020]{passino_2020_bayesian}
Passino, F.~S. and Heard, N.~A. (2020).
\newblock Bayesian estimation of the latent dimension and communities in stochastic blockmodels.
\newblock {\em Statistics and Computing}, 30:1291--1307.

\bibitem[Peres-Neto and Jackson, 2001]{peresneto_2001_how}
Peres-Neto, P.~R. and Jackson, D.~A. (2001).
\newblock How well do multivariate data sets match? {T}he advantages of a {P}rocrustean superimposition approach over the {M}antel test.
\newblock {\em Oecologia}, 129:169–178.

\bibitem[Pham and Sewell, 2024]{Pham_Sewell_2024}
Pham, H. T.~D. and Sewell, D.~K. (2024).
\newblock Automated detection of edge clusters via an overfitted mixture prior.
\newblock {\em Network Science}, 12(1):88–106.

\bibitem[{R Core Team}, 2024]{rcoreteam_2024_r}
{R Core Team} (2024).
\newblock {\em R: A Language and Environment for Statistical Computing}.
\newblock R Foundation for Statistical Computing, Vienna, Austria.

\bibitem[Raftery et~al., 2012]{raftery_2012_fast}
Raftery, A.~E., Niu, X., Hoff, P.~D., and Yeung, K.~Y. (2012).
\newblock Fast inference for the latent space network model using a case-control approximate likelihood.
\newblock {\em Journal of Computational and Graphical Statistics}, 21:901--919.

\bibitem[Rastelli et~al., 2016]{rastelli_2016_properties}
Rastelli, R., Friel, N., and Raftery, A.~E. (2016).
\newblock Properties of latent variable network models.
\newblock {\em Network Science}, 4:407--432.

\bibitem[Rastelli et~al., 2018]{rastelli_2018_computationally}
Rastelli, R., Maire, F., and Friel, N. (2018).
\newblock Computationally efficient inference for latent position network models.

\bibitem[Ro\v{c}kov\'a and George, 2016]{rokov_2016_fast}
Ro\v{c}kov\'a, V. and George, E.~I. (2016).
\newblock Fast {B}ayesian factor analysis via automatic rotations to sparsity.
\newblock {\em Journal of the American Statistical Association}, 111:1608--1622.

\bibitem[Ryan et~al., 2017]{ryan_2017_bayesian}
Ryan, C., Wyse, J., and Friel, N. (2017).
\newblock Bayesian model selection for the latent position cluster model for social networks.
\newblock {\em Network Science}, 5:70--91.

\bibitem[Salter-Townshend and Murphy, 2013]{saltertownshend_2013_variational}
Salter-Townshend, M. and Murphy, T.~B. (2013).
\newblock Variational {B}ayesian inference for the latent position cluster model for network data.
\newblock {\em Computational Statistics \& Data Analysis}, 57:661--671.

\bibitem[Scrucca et~al., 2016]{scrucca_2016_mclust}
Scrucca, L., Fop, M., Murphy, T., B., and Raftery, Adrian, E. (2016).
\newblock mclust 5: Clustering, classification and density estimation using {G}aussian finite mixture models.
\newblock {\em The R Journal}, 8:289–317.

\bibitem[Sewell, 2020]{sewell_2020_modelbased}
Sewell, D.~K. (2020).
\newblock Model-based edge clustering.
\newblock {\em Journal of Computational and Graphical Statistics}, 30:390--405.

\bibitem[Sewell and Chen, 2017]{sewell_2017_latent}
Sewell, D.~K. and Chen, Y. (2017).
\newblock Latent space approaches to community detection in dynamic networks.
\newblock {\em Bayesian Analysis}, 12:351--377.

\bibitem[Snijders and Nowicki, 1997]{snijders_1997_estimation}
Snijders, T.~A. and Nowicki, K. (1997).
\newblock Estimation and prediction for stochastic blockmodels for graphs with latent block structure.
\newblock {\em Journal of Classification}, 14:75--100.

\bibitem[Sosa and Betancourt, 2022]{sosa_2022_a}
Sosa, J. and Betancourt, B. (2022).
\newblock A latent space model for multilayer network data.
\newblock {\em Computational Statistics \& Data Analysis}, 169:107432.

\bibitem[Sosa and Buitrago, 2021]{Sosa_Buitrago_2021}
Sosa, J. and Buitrago, L. (2021).
\newblock A review of latent space models for social networks.
\newblock {\em Revista Colombiana de Estadística}, 44(1):171–200.

\bibitem[Sosa and Rodríguez, 2021]{Sosa_Rodríguez_2021}
Sosa, J. and Rodríguez, A. (2021).
\newblock A latent space model for cognitive social structures data.
\newblock {\em Social Networks}, 65:85--97.

\bibitem[Spencer et~al., 2022]{spencer_2022_faster}
Spencer, N.~A., Junker, B.~W., and Sweet, T.~M. (2022).
\newblock Faster {MCMC} for {G}aussian latent position network models.
\newblock {\em Network Science}, 10:20--45.

\bibitem[Yang et~al., 2020a]{yang_2020_simultaneous}
Yang, C., Priebe, C.~E., Park, Y., and Marchette, D.~J. (2020a).
\newblock Simultaneous dimensionality and complexity model selection for spectral graph clustering.
\newblock {\em Journal of Computational and Graphical Statistics}, 30:422--441.

\bibitem[Yang et~al., 2020b]{yang_2020_alpha}
Yang, Y., Pati, D., and Bhattacharya, A. (2020b).
\newblock $\alpha $-variational inference with statistical guarantees.
\newblock {\em The Annals of Statistics}, 48.

\end{thebibliography}

\clearpage
\appendix \label{chap:supp-lspcm}

\clearpage
\section*{Appendix}
\section{Hyperparameter specifications}
\label{app:hyperparam}

\begin{table}[htbp]
\centering
\caption{Hyperparameter specifications for the LSPCM model.}
\scalebox{.75}{
\begin{tabular}{llll}
Parameter & Hyperparam.                     & Values                               & Comments                                                                                                \\
$p_0$          &                                     & Generally 5                          &                                                                                                     \\
G          &                                     & Generally 20                         &                                                                                                     \\
$\alpha$   & $(\mu_{\alpha}, \sigma^2_{\alpha})$ & (0, 4)                               &                                                                                                     \\
$\nu$   & $(a_\nu, G b_\nu)$                               & Generally Gamma  & Increasing~$b_\nu$ increases the likelihood of a smaller number of clusters. \\ && ($a_\nu = 5, G b_\nu = G5$)                              \\
$\psi_g$   & ($a_{\psi}, b_{\psi}$)              & (400,400)   & Require expected value near 1 with low variance to avoid \\ &&& very variable clusters.                            \\
$\delta_1$ & $(a_1,b_1)$                         & (2,1)                                & Based on the work from \cite{durante_2017_a, gwee_2022_a}.                     \\
$\delta_h$ & $(a_2,b_2,t_2)$                     & (3,1,1)                              & Based on the work from \cite{durante_2017_a, gwee_2022_a}.                       \\
           & $\xi$                               & 9                                    & Based on \cite{ryan_2017_bayesian}.                                              \\
           & $\kappa_0$                          & Generally~$\kappa_0=4$               & Increasing~$\kappa_0$~decreases adaptation.                                                          \\
           & $\kappa_1$                          & Generally~$\kappa_1=5\times 10^{-4}$ & Increasing~$\kappa_1$~decreases adaptation.                                                         \\
           & $\epsilon_1$                        & Generally~$\epsilon_1=0.8$           & Increasing~$\epsilon_1$~makes it harder to decrease the number of dimensions.                                          \\
           & $\epsilon_2$                        & Generally~$\epsilon_2=0.9$           & Increasing~$\epsilon_2$~(up to ~$\epsilon_2=1$~) makes it harder to increase \\ &&& the number of dimensions.  \\
           & $\epsilon_3$                        & Generally~$\epsilon_3=5$             & Increasing~$\epsilon_3$ makes it harder to increase the number of dimensions.             
\end{tabular}
}
\end{table}

\section{Notation and terminology}
\label{app:notation}

$n$: Number of nodes. \\
$p$: The truncation dimension in the fitted LSPM.\\
$\hat{p}$: The number of effective dimensions.\\
$p^*$: The true effective dimension of the latent space.\\
$p_0$: The initial truncation level of the number of dimensions. 
\\
$\hat{p}_m$: The modal number of effective dimensions. \\
$G$: The number of mixture components. 
\\
$G_+$: The number of non-empty mixture components. \\
$G^*$: The true number of clusters.\\
$G_m$: The posterior modal number of non-empty components. \\
$\mathbf{Y}$: $n \times n$ network adjacency matrix. \\
$y_{i,j}$: Value of an edge between node $i$ and node $j$. \\
$\mathbf{Z}$: $n \times p$ matrix of latent positions.  \\
$z_{i\ell}$: The latent position of node $i$ in dimension $\ell$. \\
$q_{i,j}$: Probability of forming an edge between node $i$ and node $j$.  \\
$\alpha$: Global parameter that captures the overall connectivity level in the network.  \\
$\tau_g$: The mixing weight of component $g$. \\
$\psi_g$: The cluster $g$ specific variance scaling parameter. \\
$\nu$: The hyperparameter for $\tau_g$. 
\\
$c_{ig}$: A binary indicator variable of membership of node $i$ in component $g$. \\
$\bm{\mu}$: $G \times p$ matrix of the mean latent positions. \\ 
$\bm{\mu}_g$: The mean latent position parameter for component $g$. \\ 
$\mathbf{\Omega}$: $p \times p$ precision matrix of the latent positions. \\
$\omega_{\ell}$: Precision/global shrinkage parameter for dimension $\ell$. \\
$\delta_{h}$: Shrinkage strength from dimension $h$.  \\
$\xi$: Scaling factor in the prior covariance of the component means. 
\\
$\bm{\Theta}$: A collective term for $\bm{\tau}, \bm{\mu}, \bm{\Omega}$. \\
$a_1$: Shape parameter of gamma distribution for $\delta_1$ in dimension $1$. 
\\
$b_1$: Rate parameter of the gamma distribution for $\delta_1$ in dimension $1$. 
\\
$a_2$: Shape parameter of gamma distribution for $\delta_h$ in dimension $h$. 
\\
$b_2$: Rate parameter of the gamma distribution for $\delta_h$ in dimension $h$. 
\\
$t_2$: Left truncation point of the gamma distribution for $\delta_h$. 
\\
$a_{\nu}$: Shape parameter of gamma distribution for $\nu$. 
\\
$b_{\nu}$: Rate parameter of the gamma distribution for $\nu$. 
\\
$a_{\psi}$: Shape parameter of gamma distribution for $\psi$. 
\\
$b_{\psi}$: Rate parameter of the gamma distribution for $\psi$. 
\\
$\kappa_0$: The parameter that affects the horizontal translation of the dimension adaptation probability exponential function. 
\\
$\kappa_1$: The parameter that affects the horizontal scaling of the dimension adaptation probability exponential function. 
\\
$\epsilon_1$: The threshold for decreasing dimensions in the dimension adaptation step. 
\\
$\epsilon_2$: The threshold for adding a dimension in the dimension adaptation step. 
\\
$\epsilon_3$: The threshold for adding a dimension in the dimension adaptation step when the truncation dimension is 1. 

\clearpage
\section{Derivation of full conditional distributions}
\label{app:fullconditionals}
Links between nodes $i$ and $j$ are assumed to form probabilistically from a Bernoulli distribution:
\begin{equation}
\begin{aligned}
\notag
    y_{i,j} &\sim \text{Bin}(q_{i,j}) \\
\end{aligned}
\end{equation}
For a binary network, the probability $q_{i,j}$ is expressed in terms of a logistic model i.e.
$$\log \frac{q_{i,j}}{1-q_{i,j}} = \alpha - \Vert \mathbf{z}_i-\mathbf{z}_j \Vert^2_2.$$
Denoting $ \eta_{i,j} = \alpha - \Vert \mathbf{z}_i-\mathbf{z}_j \Vert^2_2 $, then $ q_{i,j} = \frac{\exp(\eta_{i,j})}{1+\exp(\eta_{i,j})}. \\$
The likelihood function of the LSPCM is
\begin{equation}
\begin{aligned}
\notag
    L(\mathbf{Y}|\mathbf{Z},\alpha) &= \prod_{i \neq j} P(y_{i,j} | \mathbf{z}_i, \mathbf{z}_j, \alpha)  \\
                &= \prod_{i \neq j} 
                \left[ \frac{\exp(\eta_{i,j})}{1+\exp(\eta_{i,j})} \right]^{y_{i,j}}
                \left[1 - \frac{\exp(\eta_{i,j})}{1+\exp(\eta_{i,j})} \right]^{1 - y_{i,j}} \\
                &= \prod_{i \neq j} 
                 \frac{\exp(\eta_{i,j} y_{i,j})}{1+\exp(\eta_{i,j})}, \\
\end{aligned}
\end{equation}
The joint posterior distribution of the LSPM is
\begin{equation}
\begin{aligned}  
\notag
\mathbb{P}(\alpha, \mathbf{Z}, \mathbf{C}, \bm{\tau}, \bm{\Theta} \mid \mathbf{Y}) &\propto  \mathbb{P}(\mathbf{Y} \mid \alpha, \mathbf{Z}) P(\alpha) \mathbb{P}(\mathbf{Z} \mid \bm{C}, \bm{\tau}, \bm{\Theta}) \mathbb{P}(\bm{C} \mid \bm{\tau}) \mathbb{P}(\bm{\tau}), \mathbb{P}(\bm{\Theta}) \\
\\
\\
\\
\\
\\
\\
\\
\\
\\
\end{aligned}
\end{equation}
Then
\begin{equation}
\begin{aligned}  
\notag
\mathbb{P}(\alpha, \mathbf{Z}, \mathbf{C}, \bm{\tau}, \bm{\Theta} \mid \mathbf{Y})
    &\propto 
        \left\{\prod_{i \neq j} \left[ \frac{\exp(\eta_{i,j} y_{i,j})}{1+\exp(\eta_{i,j})} \right] \right\} \\
    &\qquad \times \left\{\frac{1}{\sqrt{2\pi\sigma_{\alpha}^2}}
        \exp\left[-\frac{1}{2\sigma_{\alpha}^2}(\alpha-\mu_{\alpha})^2\right] \right\}  \\
    &\qquad \times \prod_{i=1}^n \prod_{g=1}^G \left\{ \left(\frac{1}{2\pi}\right)^{\frac{p}{2}} 
        [\text{det } (\psi_g \mathbf{\Omega})]^{\frac{1}{2}} 
    \right. \\
    & \qquad \qquad \left. \exp\left[-\frac{1}{2} (\mathbf{z}_{i} - \bm{\mu}_g)^T 
       (\psi_g \mathbf{\Omega}) (\mathbf{z}_{i} - \bm{\mu}_g)\right] \right\}^{c_{ig}} \\
    &\qquad \times \prod_{i=1}^n  \left[ \frac{n!}{c_{i1}! \dotsm c_{iG}! } \tau^{c_{i1}}_{1} \dotsm \tau^{c_{iG}}_{G}\right]\\
    &\qquad \times \left[\frac{1}{\beta(\nu)} \prod_{g=1}^G \tau_g^{\nu -1} \right]\\
    &\qquad \times \left\{ \prod_{g=1}^G 
    \left(\frac{1}{2\pi}\right)^{\frac{p}{2}} 
        [\text{det } (\xi^{-1}\mathbf{\Omega} )]^{-\frac{1}{2}}
       \right. \\
    & \qquad \qquad \left. \exp\left[-\frac{1}{2}(\bm{\mu}_g-\bm{0})^T (\xi^{-1}\mathbf{\Omega}) (\bm{\mu}_g-\bm{0})\right] \right\}\\
    &\qquad  
    \times \left\{\frac{b_{\psi}^{a_{\psi}}}{\Gamma(a_{\psi})} (\psi_g)^{a_{\psi}-1} \exp\left[-b_{\psi}(\psi_g)\right] \right\} \\
    &\qquad  
    \times \left\{\frac{b_{\nu}^{a_{\nu}}}{\Gamma(a_{\nu})} (\nu)^{a_{\nu}-1} \exp\left[-b_{\nu}(\nu)\right] \right\} \\
    &\qquad  
    \times \left\{\frac{b_1^{a_1}}{\Gamma(a_1)} (\delta_{1})^{a_1-1} \exp\left[-b_1(\delta_{1})\right] \right\} \\
    &\qquad  
    \times \left\{\prod_{h=2}^p 
        \frac{b_2^{a_2}}{\Gamma(a_2)} (\delta_{h})^{a_2-1} \exp\left[-b_2(\delta_{h})\right] \right\} 
\\ \\ \\ \\ \\ \\ \\ \\ \\ \\ \\ \\ 
\end{aligned}
\end{equation}
Indicating with $-$ the conditioning on all the remaining variables, the full conditional distribution for $\alpha$ is:
\begin{equation}
\begin{aligned}  
\notag
\mathbb{P}(\alpha \mid -)   &\propto \left\{\prod_{i \neq j} 
                 \left[ \frac{\exp(\eta_{i,j} y_{i,j})}{1+\exp(\eta_{i,j})} \right] \right\} 
                 \times \frac{1}{\sqrt{2\pi\sigma_{\alpha}^2}} \exp\left[-\frac{1}{2}\frac{(\alpha-\mu_{\alpha})^2}{\sigma_{\alpha}^2}\right] \\
                 &\propto \left\{\prod_{i \neq j} 
                 \left[ \frac{\exp(\eta_{i,j} y_{i,j})}{1+\exp(\eta_{i,j})} \right] \right\}
                 \times \exp\left[-\frac{1}{2}\frac{(\alpha-\mu_{\alpha})^2}{\sigma_{\alpha}^2}\right] \\
\log \mathbb{P}(\alpha \mid -)&\propto \sum_{i \neq j} \hspace{2pt} \left\{ \eta_{i,j} y_{i,j} - 
                    \log \left[ 1 + \exp(\eta_{i,j})\right] \right\}
                    - \frac{1}{2}\frac{(\alpha-\mu_{\alpha})^2}{\sigma_{\alpha}^2} \\
                &\propto \sum_{i \neq j} \hspace{2pt} \left\{ \eta_{i,j} y_{i,j} - 
                    \log \left[ 1 + \exp(\eta_{i,j})\right] \right\} - (\alpha-\mu_{\alpha})^2 \\
                &\propto \sum_{i \neq j} \hspace{2pt} \left\{ (\alpha - \Vert \mathbf{z}_i-\mathbf{z}_j \Vert^2_2) y_{i,j} - \log \left[  1 + \exp(\alpha - \Vert \mathbf{z}_i-\mathbf{z}_j \Vert^2_2) \right] \right\} - (\alpha-\mu_{\alpha})^2 \\
\log \mathbb{P}(\alpha \mid -)&\propto \sum_{i \neq j} \hspace{2pt} \left\{ \alpha y_{i,j} - 
                    \log \left[ 1 + \exp(\alpha - \Vert \mathbf{z}_i-\mathbf{z}_j \Vert^2_2)\right] \right\} - (\alpha-\mu_{\alpha})^2 
\end{aligned}
\end{equation}
As this is not a recognisable distribution, the Metropolis-Hasting algorithm is employed.

\vspace{15pt}
The full conditional distribution for $\mathbf{Z}$ is:
\begin{equation}
\begin{aligned}  
\notag
\mathbb{P}(\mathbf{Z} \mid -)        
    &\propto \left\{\prod_{i \neq j} 
                 \left[ \frac{\exp(\eta_{i,j} y_{i,j})}{1+\exp(\eta_{i,j})} \right] \right\} \\
                 &\qquad \times 
         \prod_{i=1}^n \prod_{g=1}^G\left\{ \left(\frac{1}{2\pi}\right)^{\frac{p}{2}} 
        [\text{det } (\psi_g\mathbf{\Omega})]^{\frac{1}{2}} 
        \exp\left[-\frac{1}{2} (\mathbf{z}_{i} - \bm{\mu}_g)^T (\psi_g
        \mathbf{\Omega}) (\mathbf{z}_{i} - \bm{\mu}_g)\right] \right\}^{c_{ig}}   \\
\log \mathbb{P}(\mathbf{Z} \mid -)     &\propto \sum_{i \neq j} \hspace{2pt} \left\{ \eta_{i,j} y_{i,j} - 
                    \log \left[ 1 + \exp(\eta_{i,j}) \right] \right\} 
                    - \sum_{i =1}^n \sum_{g=1}^G \left[\frac{1}{2} (\mathbf{z}_{i} - \bm{\mu}_g)^T  (\psi_g 
        \mathbf{\Omega}) (\mathbf{z}_{i} - \bm{\mu}_g)\right]^{c_{ig}} \\
                &\propto \sum_{i \neq j} \hspace{2pt} \left\{ \Vert \mathbf{z}_i-\mathbf{z}_j \Vert^2_2 y_{i,j} 
                - \log \left[ 1 + \exp(\alpha - \Vert \mathbf{z}_i-\mathbf{z}_j \Vert^2_2\right)] \right\} \\
                &\qquad - \sum_{i =1}^n \sum_{g=1}^G  \left[\frac{1}{2} (\mathbf{z}_{i} - \bm{\mu}_g)^T 
        (\psi_g \mathbf{\Omega}) (\mathbf{z}_{i} - \bm{\mu}_g)\right]^{c_{ig}} 
        \\
        \\
\end{aligned}
\end{equation}
As this is not a recognisable distribution, the Metropolis-Hasting algorithm is employed.

\vspace{15pt}
The full conditional distribution for $\bm{c}_{ig} = 1$ is:
\begin{equation}
\begin{aligned}  
\notag
\mathbb{P}(\bm{c}_{ig} =1 \mid -)  &= \frac{\mathbb{P}(- \mid \bm{c}_{ig} =1) \mathbb{P}(\bm{c}_{ig} =1)}{ \mathbb{P}(-)} = \frac{\mathbb{P}(\mathbf{z}_{i} \mid \bm{c}_{ig} =1) \mathbb{P}(\bm{c}_{ig} =1)}{\sum_{r=1}^G \mathbb{P}(\mathbf{z}_{ir} \mid \bm{c}_{ir} =1) \mathbb{P}(\bm{c}_{ir} =1)} \\
     &= \frac{ \text{MVN}_p(\mathbf{z}_{i} ; \bm{\mu}_g, \psi_g^{-1} \bm{\Omega}^{-1}) \tau_g}{\sum_{r=1}^G  \text{MVN}_p(\mathbf{z}_{ir} ; \bm{\mu}_r, \psi_g^{-1}\bm{\Omega}^{-1}) \tau_r}= \frac{\tau_g \text{MVN}_p(\mathbf{z}_{i} ; \bm{\mu}_g, \psi_g^{-1} \bm{\Omega}^{-1})}{\sum_{r=1}^G \tau_r \text{MVN}_p(\mathbf{z}_{ir} ; \bm{\mu}_r, \psi_g^{-1} \bm{\Omega}^{-1})  }
\end{aligned}
\end{equation}

\vspace{15pt}
The full conditional distribution for $\tau_g$ is:
\begin{equation}
\begin{aligned}  
\notag
\mathbb{P}(\tau_g \mid -)   &\propto  \left[\frac{1}{\beta(\nu)}  \tau_g^{\nu -1} \right] \left[ \prod_{i=1}^n   \tau_{g}^{\bm{c}_{ig}}\right] \\
&\propto  \tau_g^{\nu -1}    \tau_{g}^{\sum_{i=1}^n \bm{c}_{ig}} \\
&\propto  \tau_g^{\sum_{i=1}^n \bm{c}_{ig} + \nu -1}   \\
&\sim \text{Dir}(\sum_{i=1}^n \bm{c}_{ig} + \nu) 
\end{aligned}
\end{equation}

The full conditional distribution for $\nu$ is:
\begin{equation}
\begin{aligned}  
\notag
\mathbb{P}(\nu \mid -)   &\propto  \left[\frac{1}{\beta(\nu)}  \prod_{g=1}^G \tau_g^{\nu -1} \right] 
\left\{\frac{b_{\nu}^{a_{\nu}}}{\Gamma(a_{\nu})} (\nu)^{a_{\nu}-1} \exp\left[-b_{\nu}(\nu)\right] \right\}  \\
&\propto  \prod_{g=1}^G \tau_g^{\nu -1} \nu^{a_{\nu}-1} \exp\left[-b_{\nu}(\nu)\right]  
\end{aligned}
\end{equation}
As this is not a recognisable distribution, the Metropolis-Hasting algorithm is employed.

The full conditional distribution for $\psi_g$ is:
\begin{equation}
\begin{aligned}  
\notag
\mathbb{P}(\psi_g \mid -)   &\propto   \prod_{i=1}^n \left\{ \left(\frac{1}{2\pi}\right)^{\frac{p}{2}} 
        [\text{det } (\psi_g \mathbf{\Omega})]^{\frac{1}{2}} 
        \exp\left[-\frac{1}{2} (\mathbf{z}_{i} - \bm{\mu}_g)^T (\psi_g
        \mathbf{\Omega}) (\mathbf{z}_{i} - \bm{\mu}_g)\right] \right\}^{c_{ig}} \\ &\qquad \times \left\{\frac{b_{\psi}^{a_{\psi}}}{\Gamma(a_{\psi})} (\psi_g)^{a_{\psi}-1} \exp\left[-b_{\psi}(\psi_g)\right] \right\}  \\ 
        &\propto  \psi_g^{\frac{p\sum_{i=1}^n c_{ig}}{2} } 
        \exp\left[-\frac{1}{2} \sum_{i=1}^n (\mathbf{z}_{i} - \bm{\mu}_g)^T (\psi_g
        \mathbf{\Omega}) (\mathbf{z}_{i} - \bm{\mu}_g)c_{ig}\right]  \\ &\qquad \times \left\{\frac{b_{\psi}^{a_{\psi}}}{\Gamma(a_{\psi})} (\psi_g)^{a_{\psi}-1} \exp\left[-b_{\psi}(\psi_g)\right] \right\}  \\ 
        &\propto  \psi_g^{\frac{p\sum_{i=1}^n c_{ig}}{2} } (\psi_g)^{a_{\psi}-1}
        \exp\left[-\frac{1}{2} \sum_{i=1}^n (\mathbf{z}_{i} - \bm{\mu}_g)^T (\psi_g
        \mathbf{\Omega}) (\mathbf{z}_{i} - \bm{\mu}_g)c_{ig}\right]   \exp\left[-b_{\psi}(\psi_g)\right] \\ 
        &\propto  \psi_g^{a_{\psi}+\frac{p\sum_{i=1}^n c_{ig}}{2}-1 } 
        \exp\left[-\frac{1}{2} \sum_{i=1}^n (\mathbf{z}_{i} - \bm{\mu}_g)^T (\psi_g
        \mathbf{\Omega}) (\mathbf{z}_{i} - \bm{\mu}_g)c_{ig} -b_{\psi}(\psi_g)\right]   \\ 
        &\propto  \psi_g^{a_{\psi}+\frac{p\sum_{i=1}^n c_{ig}}{2}-1 } 
        \exp\left\{-\psi_g \left[b_{\psi}+ \frac{1}{2} \sum_{i=1}^n (\mathbf{z}_{i} - \bm{\mu}_g)^T (
        \mathbf{\Omega}) (\mathbf{z}_{i} - \bm{\mu}_g)c_{ig} \right]\right\}  \\
        &\sim \text{Gam} \left( a_{\psi}+\frac{p\sum_{i=1}^n c_{ig}}{2} \quad,  \quad b_{\psi}+ \frac{1}{2} \sum_{i=1}^n  (\mathbf{z}_{i} - \bm{\mu}_g)^T (
        \mathbf{\Omega}) (\mathbf{z}_{i} - \bm{\mu}_g)c_{ig} \right) \\
\end{aligned}
\end{equation}

The full conditional distribution for $\bm{\mu}_g$ is:
\begin{equation}
\begin{aligned}  
\notag
\mathbb{P}(\bm{\mu}_g \mid -)        
    &\propto \prod_{i=1}^n \left\{ \left(\frac{1}{2\pi}\right)^{\frac{p}{2}} 
        [\text{det } (\psi_g\mathbf{\Omega})]^{\frac{1}{2}} 
        \exp\left[-\frac{1}{2} (\mathbf{z}_{i} - \bm{\mu}_g)^T 
        (\psi_g\mathbf{\Omega}) (\mathbf{z}_{i} - \bm{\mu}_g)\right] \right\}^{c_{ig}} \\
    &\qquad \times \left\{ \left(\frac{1}{2\pi}\right)^{\frac{p}{2}} 
        [\text{det } (\xi^{-1}\mathbf{\Omega})]^{\frac{1}{2}}
        \exp\left[-\frac{1}{2}(\bm{\mu}_g-\bm{0})^T (\xi^{-1}\bm{\Omega}) (\bm{\mu}_g-\bm{0})\right] \right\}\\
    &\propto \exp\left\{ \sum_{i=1}^n  \left[ -\frac{1}{2} (\mathbf{z}_{i} - \bm{\mu}_g)^T 
        (\psi_g\mathbf{\Omega}) (\mathbf{z}_{i} - \bm{\mu}_g) (c_{ig}) \right] -\frac{\bm{\mu}_{g}^T (\xi^{-1}\bm{\Omega})\bm{\mu}_{g}}{2} \right\} \\
    &\propto \exp\left\{-\frac{1}{2} \left[ \sum_{i=1}^n  \left[  (\mathbf{z}_{i}^T\psi_g\bm{\Omega}\mathbf{z}_{i} c_{ig} - 2\mathbf{z}_{i}^T\psi_g\bm{\Omega} \bm{\mu}_{g} c_{ig} + \bm{\mu}_{g}^T \psi_g\bm{\Omega} \bm{\mu}_{g} c_{ig}) \right] + \bm{\mu}_{g}^T\xi^{-1}\bm{\Omega}\bm{\mu}_{g} \right] \right\} \\
    &\propto \exp\left\{-\frac{1}{2} \left[  - 2 \sum_{i=1}^n \psi_g c_{ig} \mathbf{z}_{i}^T \bm{\Omega} \bm{\mu}_{g} + \sum_{i=1}^n \psi_g c_{ig}  \bm{\mu}_{g}^T \bm{\Omega} \bm{\mu}_{g}  + \bm{\mu}_{g}^T\xi^{-1}\bm{\Omega}\bm{\mu}_{g} \right] \right\} \\
    &\propto \exp\left\{-\frac{1}{2} \left[ \bm{\mu}_{g}^T \bm{\Omega} \left(\psi_g\sum_{i=1}^n c_{ig}  + \xi^{-1}\right)
    \bm{\mu}_{g}  - 2 \bm{\mu}_{g}^T
    \left(\psi_g\bm{\Omega} \sum_{i=1}^n (c_{ig}  \mathbf{z}_{i}) \right) \right] \right\} \\
    &\propto \exp\left\{-\frac{1}{2} \left[ \bm{\mu}_{g}^T - \left\{\psi_g\bm{\Omega}  \sum_{i=1}^n (c_{ig}  \mathbf{z}_{i}) \right\}^T \left\{ \bm{\Omega} \left(\psi_g\sum_{i=1}^n c_{ig}  +\xi^{-1}\right) \right \}^{-1} \right]  \right. \\
    &\quad\times \left. \left[ \bm{\Omega} \left(\psi_g \sum_{i=1}^n c_{ig}  + \xi^{-1} \right) \right]\left[ \bm{\mu}_{g} - \left\{ \bm{\Omega}^T \left(\psi_g\sum_{i=1}^n c_{ig}  +\xi^{-1}\right) \right \}^{-1}  \left\{\psi_g\bm{\Omega}  \sum_{i=1}^n (c_{ig}  \mathbf{z}_{i}) \right\}\right]\right\} \\ 
\text{since } &\bm{\Omega} \text{ is a diagonal matrix, } \bm{\Omega} = \bm{\Omega}^T \text{ and }\bm{\Omega}\bm{\Omega}^{-1} = \mathbf{I} \text{, thus,} \\
    &\propto \exp\left\{-\frac{1}{2} \left[ \bm{\mu}_{g} - \frac{ \sum_{i=1}^n c_{ig}  \mathbf{z}_{i}}{ \sum_{i=1}^n c_{ig}  +\xi^{-1} } \right]^T \left[ \bm{\Omega} \left(\psi_g\sum_{i=1}^n c_{ig}  + \xi^{-1}\right)  \right] \right. \\
    &\qquad\times \left. \left[ \bm{\mu}_{g} - \frac{ \sum_{i=1}^n c_{ig}  \mathbf{z}_{i}}{ \sum_{i=1}^n c_{ig}  + \xi^{-1} } \right]\right\} \\
    &\sim \text{MVN}_p\left(\frac{\sum_{i=1}^n c_{ig}  \mathbf{z}_{i}}{ \sum_{i=1}^n c_{ig}  + \xi^{-1} } \quad,\quad \left[ \bm{\Omega} \left(\psi_g\sum_{i=1}^n c_{ig}  + \xi^{-1} \right) \right]^{-1} \right) 
    \\
    \\
    \\
    \\
    \\
    \\
    \\
    \\
    \\
    \\
    \\
    \\
\end{aligned}
\end{equation}

The full conditional distribution for $\delta_{1}$ is:
\begin{equation}
\begin{aligned}  
\notag
\mathbb{P}(\delta_{1} \mid -)    
&\propto \prod_{i=1}^n \prod_{g=1}^G \left\{ \left(\frac{1}{2\pi}\right)^{\frac{p}{2}} 
        [\text{det } (\psi_g \mathbf{\Omega})]^{\frac{1}{2}} 
        \exp\left[-\frac{1}{2} (\mathbf{z}_{i} - \bm{\mu}_g)^T (\psi_g
        \mathbf{\Omega}) (\mathbf{z}_{i} - \bm{\mu}_g)\right] \right\}^{c_{ig}} \\
    &\qquad \times \prod_{g=1}^G \left\{ \left(\frac{1}{2\pi}\right)^{\frac{p}{2}} 
        [\text{det } (\xi^{-1}\mathbf{\Omega})]^{\frac{1}{2}}
        \exp\left[-\frac{1}{2}(\bm{\mu}_g-\bm{0})^T (\xi^{-1}\bm{\Omega}) (\bm{\mu}_g-\bm{0})\right] \right\}\\
    &\qquad \times \left[\frac{b_1^{a_1}}{\Gamma(a_1)} (\delta_{1})^{a_1-1} \exp\left(-b_1\delta_{1}\right) \right] \\
    &\propto \left\{ 
         \bm{\omega}^{\frac{p \sum_{i=1}^{n} \sum_{i=g}^{G} c_{ig}}{2}} \mathbf{I}_p
        \exp\left[-\frac{1}{2} \sum_{i=1}^{n} \sum_{i=g}^{G}(\mathbf{z}_{i} - \bm{\mu}_g)^T 
        (\psi_g\bm{\omega}) \mathbf{I}_p(\mathbf{z}_{i} - \bm{\mu}_g) c_{ig} \right] \right\} \\
    &\qquad \times \left\{ \bm{\omega}^{\frac{Gp}{2}} \mathbf{I}_p
        \exp\left[-\frac{1}{2}\sum_{g=1}^{G}\bm{\mu}_g^T (\xi^{-1}\bm{\omega}) \mathbf{I}_p \bm{\mu}_g\right] \right\} \times \left[ \delta_{1}^{a_1-1} \exp(-b_1\delta_{1})\right]  \\
    &\propto \left\{ 
         \delta_{1}^{\frac{np}{2}} 
        \exp\left[-\frac{1}{2} \sum_{i=1}^{n} \sum_{i=g}^{G} (\mathbf{z}_{i} - \bm{\mu}_g)^T (\psi_g)
        \left(\prod_{m=1}^{\ell} \delta_{m}\right)\mathbf{I}_p (\mathbf{z}_{i} - \bm{\mu}_g) c_{ig} \right] \right\} \\
    &\qquad \times \left\{ \delta_{1}^{\frac{Gp}{2}}
        \exp\left[-\frac{1}{2}\sum_{g=1}^{G}\bm{\mu}_g^T (\xi^{-1}) \left(\prod_{m=1}^{\ell} \delta_{m}\right)\mathbf{I}_p \bm{\mu}_g\right] \right\} \times \left[ \delta_{1}^{a_1-1} \exp(-b_1\delta_{1})\right]  \\
    &\propto  
     \delta_{1}^{\frac{np}{2} } \delta_{1}^{\frac{Gp}{2}}\delta_{1}^{a_1 - 1} 
    \exp\left[-\frac{1}{2} \sum_{i=1}^{n} \sum_{i=g}^{G} (\mathbf{z}_{i} - \bm{\mu}_g)^T (\psi_g)
     \left( \delta_{1} \prod_{m=2}^{\ell} \delta_{m}\right)\mathbf{I}_p (\mathbf{z}_{i} - \bm{\mu}_g) c_{ig} \right.\\
     &\qquad \left. -\frac{1}{2}\sum_{g=1}^{G}\bm{\mu}_g^T (\xi^{-1}) \left( \delta_{1} \prod_{m=2}^{\ell} \delta_{m}\right)\mathbf{I}_p \bm{\mu}_g - b_1\delta_{1} \right]  \\
    &\propto  
     \delta_{1}^{\frac{\left(n+G\right) p}{2} + a_1 - 1} 
    \exp \left\{ -\left[\frac{1}{2} \sum_{i=1}^{n} \sum_{g=1}^{G} (\mathbf{z}_{i} - \bm{\mu}_g)^T 
    (\psi_g) \left(\prod_{m=2}^{\ell} \delta_{m}\right)\mathbf{I}_p (\mathbf{z}_i - \bm{\mu}_g) c_{ig} \right. \right. \\
    &\qquad \left. \left. +\frac{1}{2}\sum_{g=1}^{G}\bm{\mu}_g^T (\xi^{-1})\left( \prod_{m=2}^{\ell} \delta_{m}\right)\mathbf{I}_p \bm{\mu}_g +b_1 \right] \delta_{1} \right\}  \\
    &\sim \text{Gam} \left( \frac{\left( n+G\right) p}{2} + a_1 \quad,  \quad \frac{1}{2} \sum_{i=1}^{n} \sum_{g=1}^{G} (\mathbf{z}_{i} - \bm{\mu}_g)^T 
    (\psi_g) \left(\prod_{m=2}^{\ell} \delta_{m}\right)\mathbf{I}_p (\mathbf{z}_{i} - \bm{\mu}_g) c_{ig}  \right. \\
    &\left. \qquad \qquad \qquad \qquad \qquad \qquad \qquad \qquad \quad+ \frac{1}{2}\sum_{g=1}^{G}\bm{\mu}_g^T (\xi^{-1}) \left( \prod_{m=2}^{\ell} \delta_{m}\right)\mathbf{I}_p \bm{\mu}_g +b_1  \right) \\
    \\
    \\
    \\
    \\
    \\
    \\
\end{aligned}
\end{equation}

The full conditional distribution for $\delta_{h}$, where $h \geq 2$ is:
\begin{equation}
\begin{aligned}  
\notag
\mathbb{P}(\delta_{h} \mid -)      
&\propto \prod_{i=1}^n \prod_{g=1}^G \left\{ \left(\frac{1}{2\pi}\right)^{\frac{p}{2}} 
        [\text{det } (\psi_g \mathbf{\Omega})]^{\frac{1}{2}} 
        \exp\left[-\frac{1}{2} (\mathbf{z}_{i} - \bm{\mu}_g)^T (\psi_g
        \mathbf{\Omega}) (\mathbf{z}_{i} - \bm{\mu}_g)\right] \right\}^{c_{ig}} \\
    &\qquad \times \prod_{g=1}^G \left\{ \left(\frac{1}{2\pi}\right)^{\frac{p}{2}} 
        [\text{det } (\xi^{-1}\mathbf{\Omega})]^{\frac{1}{2}}
        \exp\left[-\frac{1}{2}(\bm{\mu}_g-\bm{0})^T (\xi^{-1}\bm{\Omega}) (\bm{\mu}_g-\bm{0})\right] \right\}\\
    &\qquad \times \left[
        \frac{b_2^{a_2}}{\Gamma(a_2)} (\delta_{h})^{a_2-1} \exp(-b_2\delta_{h}) \right]  \\
    &\propto \left\{ 
         \bm{\omega}^{\frac{(p-h+1) \sum_{i=1}^{n} \sum_{g=1}^{G} c_{ig}}{2}} \mathbf{I}_p
        \exp\left[-\frac{1}{2} \sum_{i=1}^{n} \sum_{g=1}^{G}(\mathbf{z}_{i} - \bm{\mu}_g)^T (\psi_g
        \bm{\omega}) \mathbf{I}_p(\mathbf{z}_{i} - \bm{\mu}_g) c_{ig} \right] \right\} \\
    &\qquad \times \left\{ \bm{\omega}^{\frac{Gp}{2}} \mathbf{I}_p
        \exp\left[-\frac{1}{2}\sum_{g=1}^{G}\bm{\mu}_g^T (\xi^{-1}\bm{\omega}) \mathbf{I}_p \bm{\mu}_g\right] \right\}\\
    &\qquad \times \left[ \delta_{h}^{a_2-1} \exp(-b_2\delta_{h}) \right] \\
    &\propto \left\{ 
         \delta_{h}^{\frac{(p-h+1) n}{2}} 
        \exp\left[-\frac{1}{2} \sum_{i=1}^{n} \sum_{g=1}^{G}(\mathbf{z}_{i} - \bm{\mu}_g)^T (\psi_g)
        \left(\prod_{m=1}^{\ell} \delta_{m}\right)\mathbf{I}_p (\mathbf{z}_{i} - \bm{\mu}_g) c_{ig} \right] \right\} \\
    &\qquad \times \left\{ \delta_{h}^{\frac{G(p-h+1)}{2}}
        \exp\left[-\frac{1}{2}\sum_{g=1}^{G}\bm{\mu}_g^T (\xi^{-1})\left(\prod_{m=1}^{\ell} \delta_{m}\right)\mathbf{I}_p \bm{\mu}_g\right] \right\}\\
    &\qquad \times \left[ \delta_{h}^{a_2-1} \exp(-b_2\delta_{h}) \right] \\
    &\propto  
     \delta_{h}^{\frac{n (p-h+1)}{2}}  \delta_{h}^{\frac{G(p-h+1)}{2}} \delta_{h}^{a_2 - 1} 
    \exp\left[-\frac{1}{2} \sum_{i=1}^{n} \sum_{g=1}^{G} (\mathbf{z}_{i} - \bm{\mu}_g)^T (\psi_g)
     \left( \delta_{h} \prod_{m=1, m \neq h}^{\ell} \delta_{m}\right) \right.\\
     &\qquad \left. \times \mathbf{I}_p (\mathbf{z}_{i} - \bm{\mu}_g) c_{ig}-\frac{1}{2}\sum_{g=1}^{G}\bm{\mu}_g^T (\xi^{-1}) \left( \delta_{h} \prod_{m=1, m \neq h}^{\ell} \delta_{m}\right)\mathbf{I}_p \bm{\mu}_g -b_2\delta_{h} \right]  \\
    &\propto  
     \delta_{h}^{\frac{\left( n+G\right)(p-h+1) }{2} + a_2 - 1} 
    \exp \left\{ -\left[\frac{1}{2} \sum_{i=1}^{n} \sum_{g=1}^{G}(\mathbf{z}_{i} - \bm{\mu}_g)^T (\psi_g)
    \left(\prod_{m=1, m \neq h}^{\ell} \delta_{m}\right)\mathbf{I}_p (\mathbf{z}_{i} - \bm{\mu}_g) c_{ig} \right. \right. \\
    &\qquad \left. \left. +\frac{1}{2}\sum_{g=1}^{G}\bm{\mu}_g^T (\xi^{-1})\left(\prod_{m=1, m \neq h}^{\ell} \delta_{m}\right)\mathbf{I}_p \bm{\mu}_g +b_2 \right] \delta_{h} \right\}  \\
\end{aligned}
\end{equation}
since $\delta_{h}$ is bounded between $[1, \infty)$,
\begin{equation}
\begin{aligned}  
\notag
    \sim \text{Gam}^\text{T} &\left( \frac{\left( n+G\right)(p-h+1) }{2} + a_2 \quad, \quad \frac{1}{2} \sum_{i=1}^{n} (\mathbf{z}_{i} - \bm{\mu}_g)^T (\psi_g)
    \left(\prod_{m=1, m \neq h}^{\ell} \delta_{m}\right)\mathbf{I}_p (\mathbf{z}_{i} - \bm{\mu}_g) c_{ig} \right. \\
    &\left. \qquad \qquad \qquad \qquad \qquad \qquad \qquad + \frac{1}{2}\sum_{g=1}^{G}\bm{\mu}_g^T (\xi^{-1})\left(\prod_{m=1, m \neq h}^{\ell} \delta_{m}\right)\mathbf{I}_p \bm{\mu}_g  +b_2 \quad , \quad 1 \right)
\end{aligned}
\end{equation}

\clearpage
\section{Simulation studies' additional posterior distribution plots}
\label{app:simstudies}

The performance of LSPCM is also assessed through the posterior distributions of latent positions' variance, shrinkage strength, and the parameter $\alpha$. Included here are supplementary plots to assist in assessing the performance of the LSPCM.

\begin{figure}[htbp]
     \centering
     \begin{subfigure}[b]{0.31\linewidth}
         \centering
         \includegraphics[width=\linewidth]{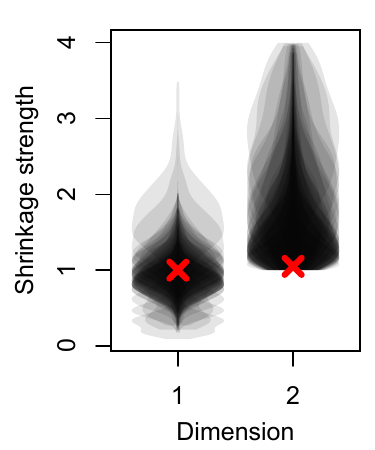}
         \caption{}
         \label{fig:n50p2g3_pmd}
     \end{subfigure}
     \hfill
     \begin{subfigure}[b]{0.31\linewidth}
         \centering
         \includegraphics[width=\linewidth]{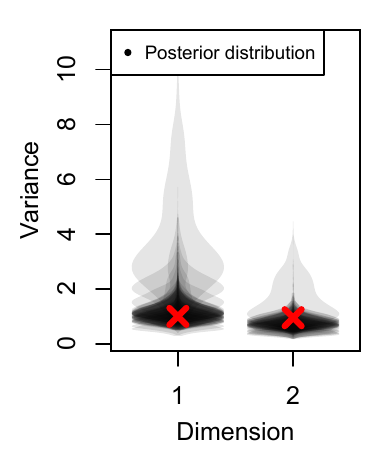}
         \caption{}
         \label{fig:n50p2g3_pmv}
     \end{subfigure}
     \begin{subfigure}[b]{0.31\linewidth}
         \centering
         \includegraphics[width=\linewidth]{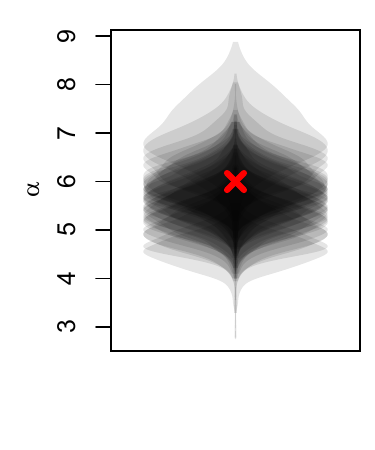}
         \caption{}
         \label{fig:n50p2g3_alpha}
     \end{subfigure}
        \caption{For the simulated network with $p^*=2$ and $G^*=3$, the posterior distribution of (a) the shrinkage strength parameter, (b) the latent position variance parameter, and (c) the $\alpha$ parameter.}
        \label{fig:n50p2g3_supp}
\end{figure}

\begin{figure}[htbp]
     \centering
     \begin{subfigure}[b]{0.31\linewidth}
         \centering
         \includegraphics[width=\linewidth]{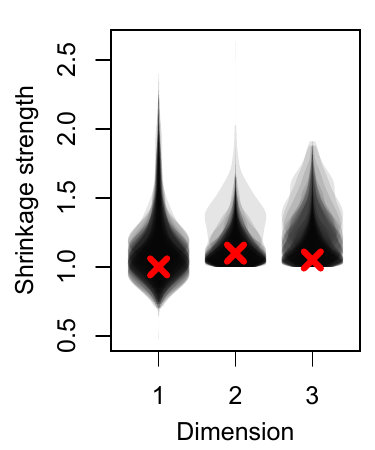}
         \caption{}
         \label{fig:n200p4g7_pmd}
     \end{subfigure}
     \hfill
     \begin{subfigure}[b]{0.31\linewidth}
         \centering
         \includegraphics[width=\linewidth]{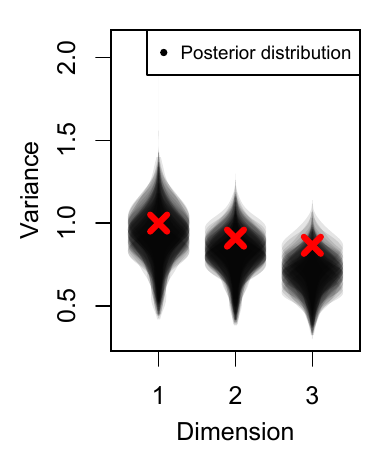}
         \caption{}
         \label{fig:n200p4g7_pmv}
     \end{subfigure}
     \begin{subfigure}[b]{0.31\linewidth}
         \centering
         \includegraphics[width=\linewidth]{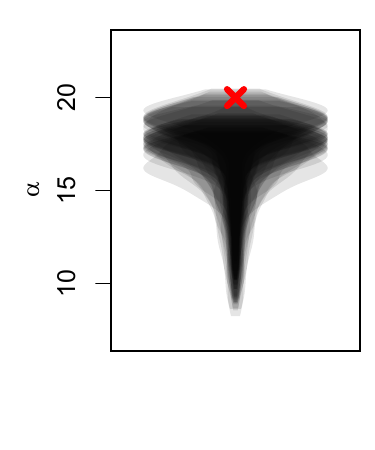}
         \caption{}
         \label{fig:n200p4g7_alpha}
     \end{subfigure}
        \caption{For the simulated network with $p^*=3$ and $G^*=7$, the posterior distribution of (a) the shrinkage strength parameter, (b) the latent position variance parameter, and (c) the $\alpha$ parameter.}
        \label{fig:n200p4g7_supp}
\end{figure}






\clearpage
\section{Additional results from analyses of Twitter networks} \label{app:application}

\begin{figure}[htbp]
     \centering
     \begin{subfigure}[b]{0.24\linewidth}
         \centering
         \includegraphics[width=\linewidth]{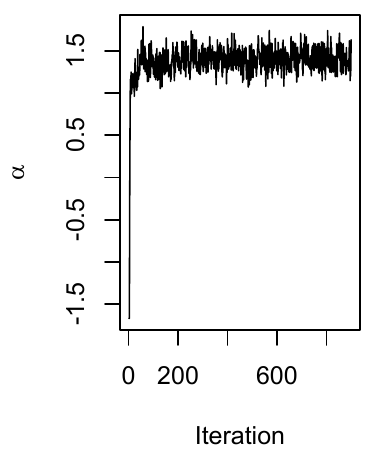}
         \caption{}
         \label{fig:football_alpha_trace}
     \end{subfigure}
     \hfill
     \begin{subfigure}[b]{0.24\linewidth}
         \centering
         \includegraphics[width=\linewidth]{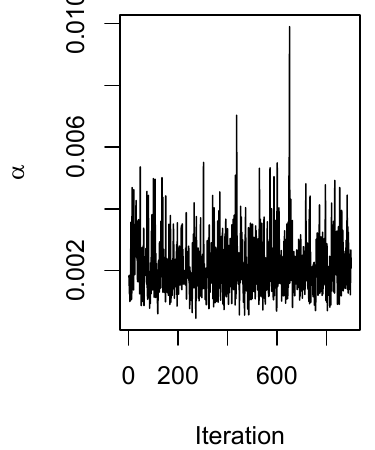}
         \caption{}
         \label{fig:football_nu_trace}
     \end{subfigure}
     \begin{subfigure}[b]{0.24\linewidth}
         \centering
         \includegraphics[width=\linewidth]{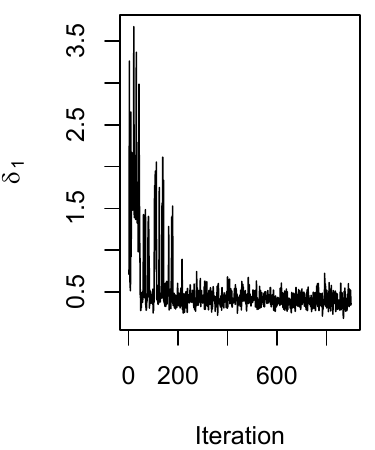}
         \caption{}
         \label{fig:football_delta1_trace}
     \end{subfigure}
     \begin{subfigure}[b]{0.24\linewidth}
         \centering
         \includegraphics[width=\linewidth]{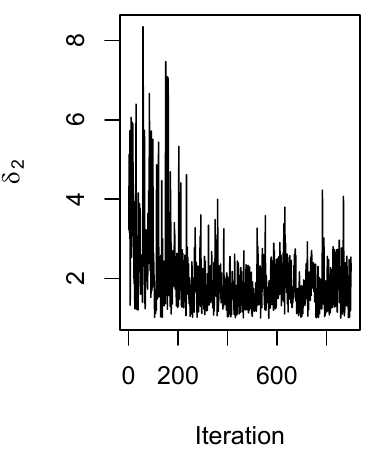}
         \caption{}
         \label{fig:football_delta2_trace}
     \end{subfigure}
        \caption{For the football Twitter network: an example of a trace plot for the parameters (a) $\alpha$, (b) $\nu$, (c) $\delta_1$, and (d) $\delta_2$.}
        \label{fig:football_trace}
\end{figure}

\begin{figure}[htbp]
     \centering
     \begin{subfigure}[b]{0.31\linewidth}
         \centering
         \includegraphics[width=\linewidth]{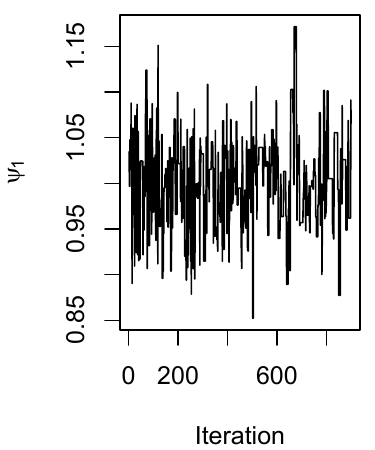}
         \caption{}
         \label{fig:football_psi1_trace}
     \end{subfigure}
     \hfill
     \begin{subfigure}[b]{0.31\linewidth}
         \centering
         \includegraphics[width=\linewidth]{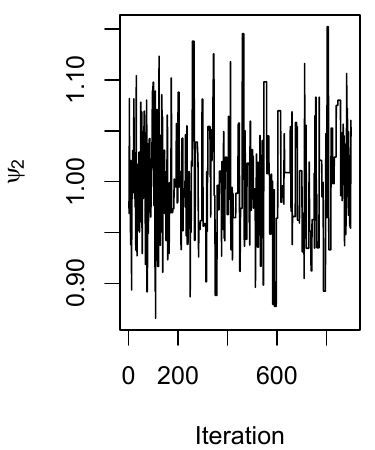}
         \caption{}
         \label{fig:football_psi2_trace}
     \end{subfigure}
     \begin{subfigure}[b]{0.31\linewidth}
         \centering
         \includegraphics[width=\linewidth]{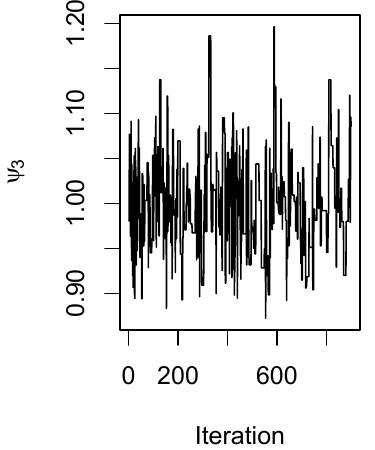}
         \caption{}
         \label{fig:football_psi3_trace}
     \end{subfigure}
        \caption{For the football Twitter network: an example of a trace plot for the parameters (a) $\psi_1$, (b) $\psi_2$ and (c) $\psi_3$.}
        \label{fig:football_psi_trace}
\end{figure}

\begin{figure}[htbp]
     \centering
     \begin{subfigure}[b]{0.32\linewidth}
         \centering
         \includegraphics[width=.95\linewidth]{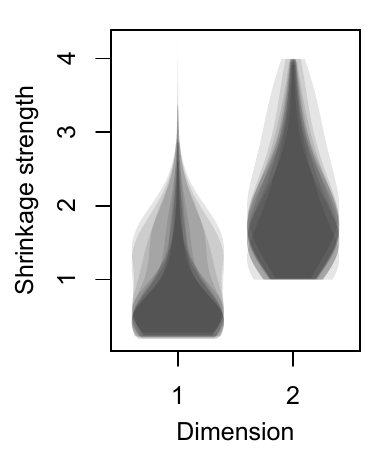}
         \caption{}
         \label{fig:football_pmd}
     \end{subfigure}
     \hfill
     \begin{subfigure}[b]{0.32\linewidth}
         \centering
         \includegraphics[width=.95\linewidth]{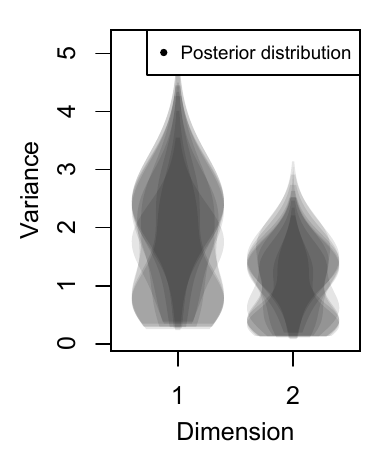}
         \caption{}
         \label{fig:football_pmv}
     \end{subfigure}
    \begin{subfigure}[b]{0.32\linewidth}
         \centering
         \includegraphics[width=.95\linewidth]{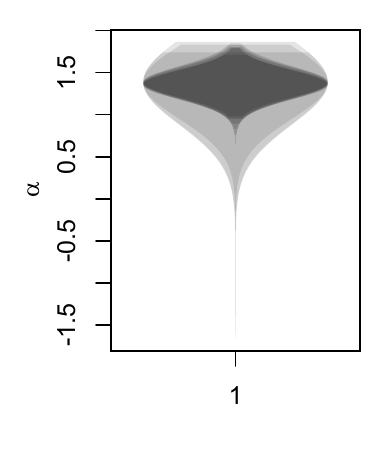}
         \caption{}
         \label{fig:football_alpha}
     \end{subfigure}
        \caption{For the football Twitter network, (a) the posterior distributions of the shrinkage strength parameters across dimensions, (b) the posterior distributions of the variance parameters across dimensions, and (c) the posterior distribution of $\alpha$.}
        \label{fig:footballviolin}
\end{figure}

\begin{figure}[htbp]
     \centering
     \begin{subfigure}[b]{0.16\linewidth}
         \centering
         \includegraphics[width=\linewidth]{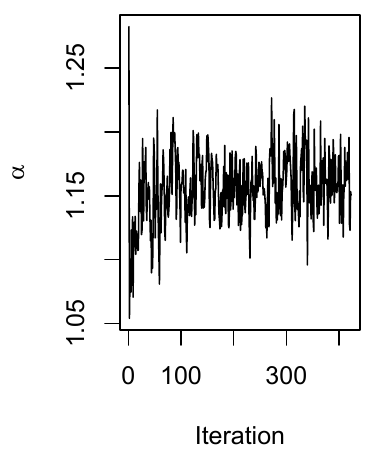}
         \caption{}
         \label{fig:irish_alpha_trace}
     \end{subfigure}
     \hfill
     \begin{subfigure}[b]{0.16\linewidth}
         \centering
         \includegraphics[width=\linewidth]{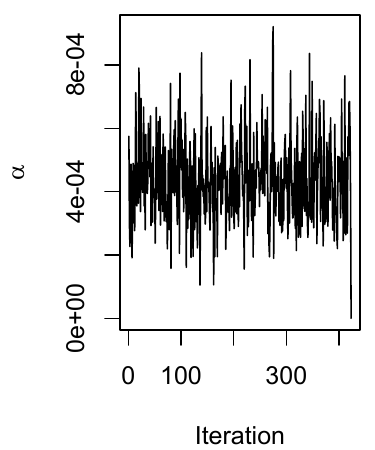}
         \caption{}
         \label{fig:irish_nu_trace}
     \end{subfigure}
     \begin{subfigure}[b]{0.16\linewidth}
         \centering
         \includegraphics[width=\linewidth]{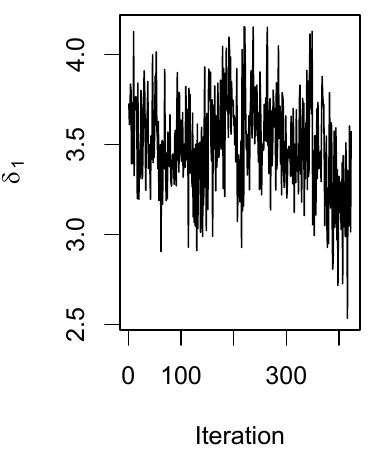}
         \caption{}
         \label{fig:irish_delta1_trace}
     \end{subfigure}
     \begin{subfigure}[b]{0.16\linewidth}
         \centering
         \includegraphics[width=\linewidth]{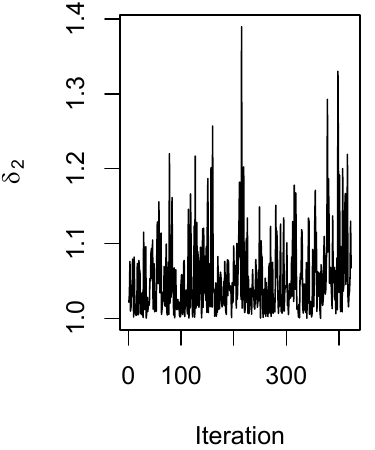}
         \caption{}
         \label{fig:irish_delta2_trace}
     \end{subfigure}
     \begin{subfigure}[b]{0.16\linewidth}
         \centering
         \includegraphics[width=\linewidth]{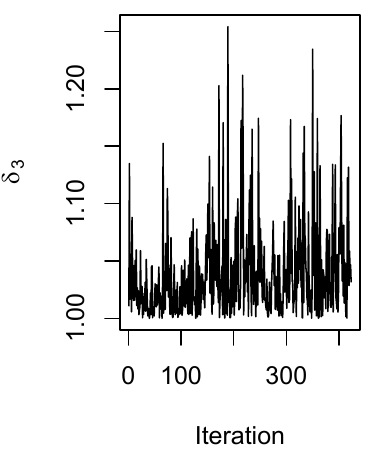}
         \caption{}
         \label{fig:irish_delta3_trace}
     \end{subfigure}
     \begin{subfigure}[b]{0.16\linewidth}
         \centering
         \includegraphics[width=\linewidth]{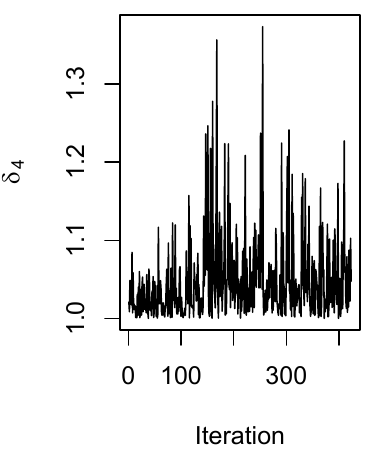}
         \caption{}
         \label{fig:irish_delta4_trace}
     \end{subfigure}
        \caption{For the Irish politicians Twitter network: an example of a trace plot for the parameters (a) $\alpha$, (b) $\nu$, (c) $\delta_1$, (d) $\delta_2$, (e) $\delta_3$, and (f) $\delta_4$.}
        \label{fig:irish_trace}
\end{figure}

\begin{figure}[htbp]
     \centering
     \begin{subfigure}[b]{0.19\linewidth}
         \centering
         \includegraphics[width=\linewidth]{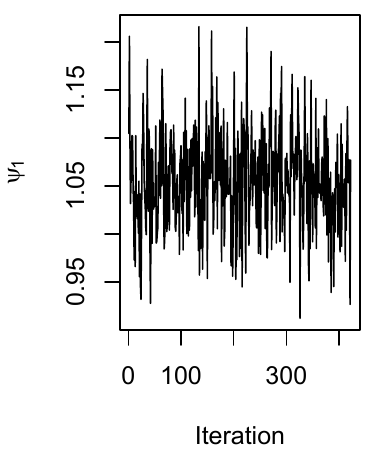}
         \caption{}
         \label{fig:irish_psi1_trace}
     \end{subfigure}
     \hfill
     \begin{subfigure}[b]{0.19\linewidth}
         \centering
         \includegraphics[width=\linewidth]{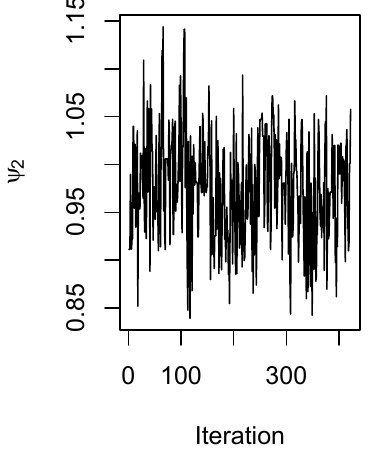}
         \caption{}
         \label{fig:irish_psi2_trace}
     \end{subfigure}
     \begin{subfigure}[b]{0.19\linewidth}
         \centering
         \includegraphics[width=\linewidth]{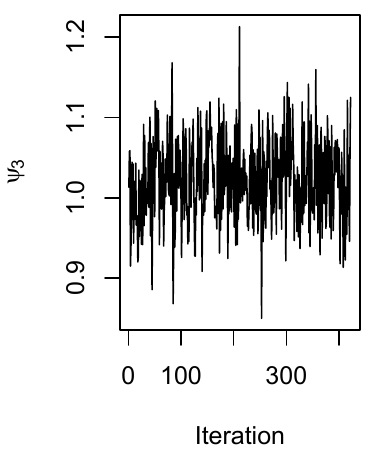}
         \caption{}
         \label{fig:irish_psi3_trace}
     \end{subfigure}
     \begin{subfigure}[b]{0.19\linewidth}
         \centering
         \includegraphics[width=\linewidth]{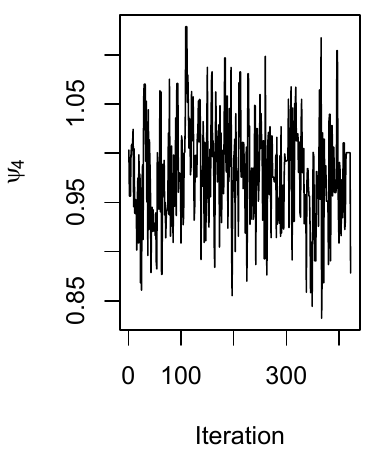}
         \caption{}
         \label{fig:irish_psi4_trace}
     \end{subfigure}
     \begin{subfigure}[b]{0.19\linewidth}
         \centering
         \includegraphics[width=\linewidth]{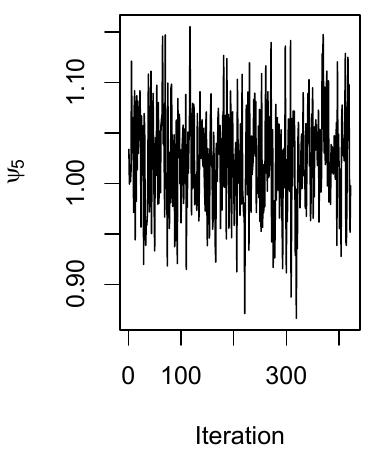}
         \caption{}
         \label{fig:irish_psi5_trace}
     \end{subfigure}
        \caption{For the Irish politicians' Twitter network: an example of a trace plot for the parameters (a) $\psi_1$, (b) $\psi_2$, (c) $\psi_3$, (d) $\psi_4$  and (e) $\psi_5$.}
        \label{fig:irish_psi_trace}
\end{figure}

\begin{figure}[htbp]
     \centering
     \begin{subfigure}[b]{0.32\linewidth}
         \centering
         \includegraphics[width=\linewidth]{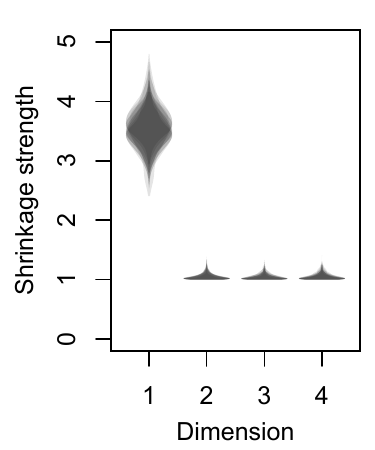}
         \caption{}
         \label{fig:ie_pmd}
     \end{subfigure}
     \hfill
     \begin{subfigure}[b]{0.32\linewidth}
         \centering
         \includegraphics[width=\linewidth]{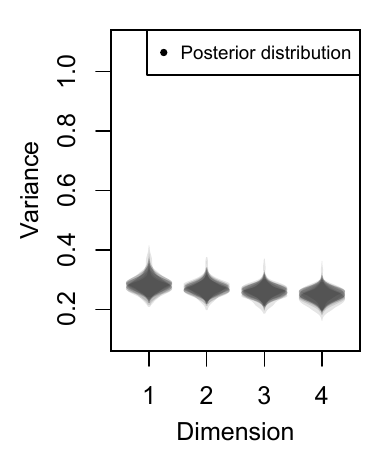}
         \caption{}
         \label{fig:ie_pmv}
     \end{subfigure}
    \begin{subfigure}[b]{0.32\linewidth}
         \centering
         \includegraphics[width=\linewidth]{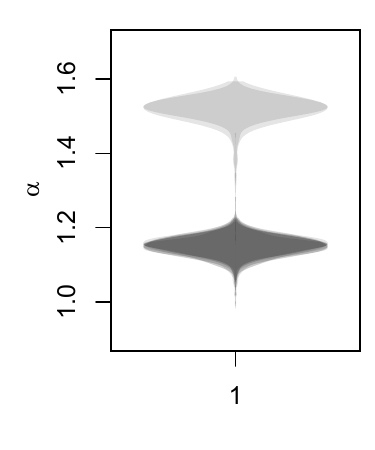}
         \caption{}
         \label{fig:ie_alpha}
     \end{subfigure}
        \caption{Irish politicians Twitter network, (a) the posterior distribution of the shrinkage strength parameters across dimensions, (b) the posterior distributions of the variance parameters across dimensions, and (c) the posterior distribution of $\alpha$.}
        \label{fig:irishviolin}
\end{figure}

\begin{figure}[htbp]
     \centering
     \begin{subfigure}[b]{0.59\linewidth}
         \centering
         \includegraphics[width=\linewidth]{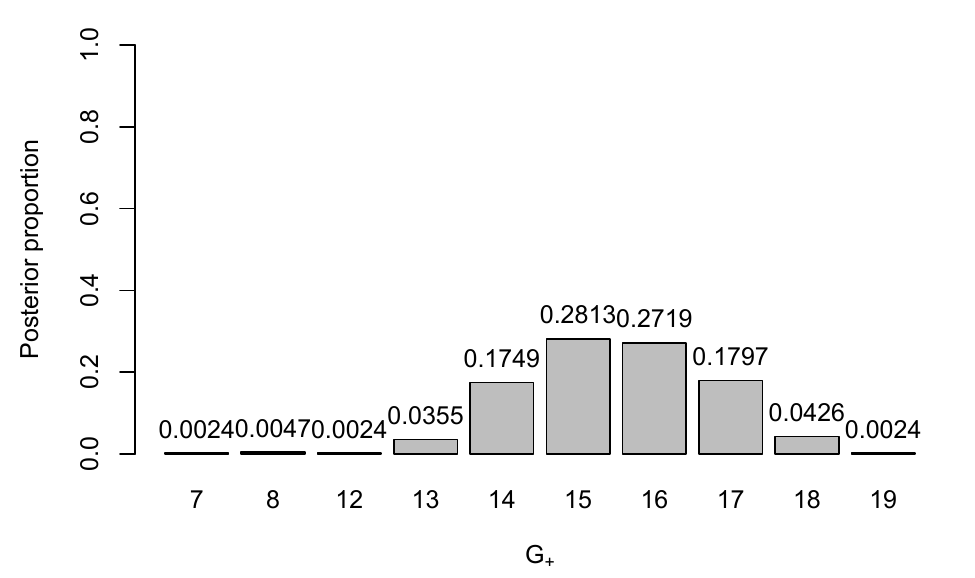}
         \caption{}
         \label{fig:irish_cluster_bar_nu_high}
     \end{subfigure}
     \hfill
     \begin{subfigure}[b]{0.39\linewidth}
         \centering
         \includegraphics[width=\linewidth]{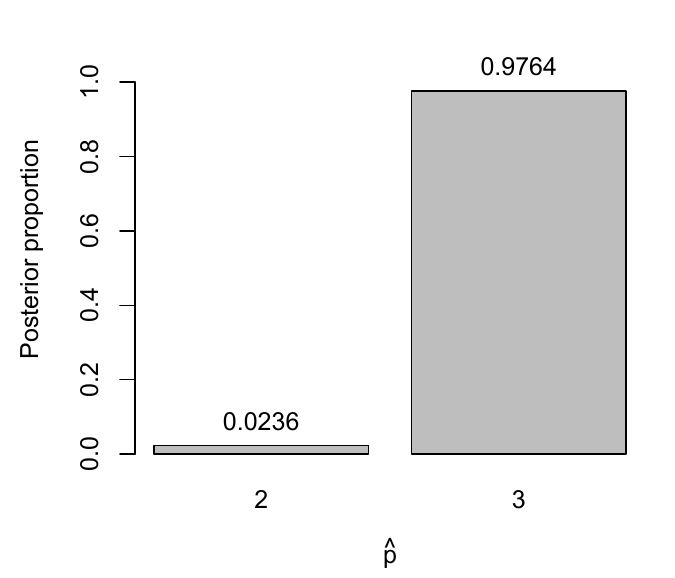}
         \caption{}
         \label{fig:irish_dim_bar_nu_high}
     \end{subfigure}
        \caption{For the Irish politicians network with $a_{\nu}=5, b_{\nu}=5G$, where $G=20$: (a) the posterior distribution of the number of non-empty components $G_+$, and (b) the distribution of the effective latent space dimension $\hat{p}$.}
        \label{fig:irish_nu_high}
\end{figure}

\begin{figure}[htbp]
     \centering
     \begin{subfigure}[b]{0.59\linewidth}
         \centering
         \includegraphics[width=\linewidth]{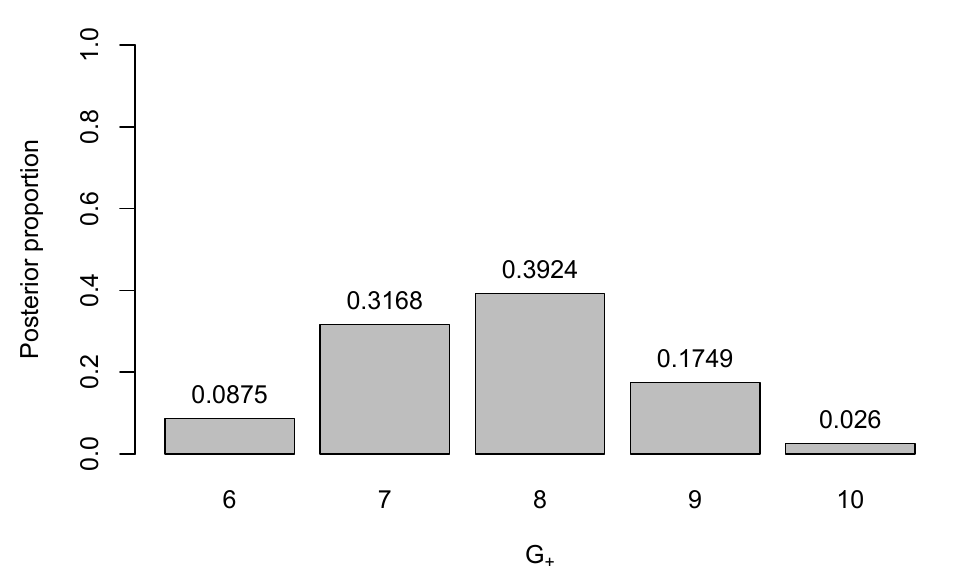}
         \caption{}
         \label{fig:irish_cluster_bar_nu_mid}
     \end{subfigure}
     \hfill
     \begin{subfigure}[b]{0.39\linewidth}
         \centering
         \includegraphics[width=\linewidth]{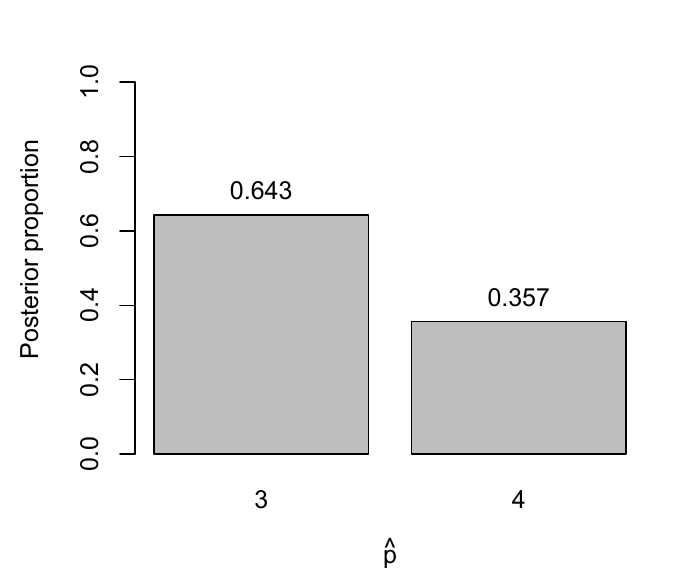}
         \caption{}
         \label{fig:irish_dim_bar_nu_mid}
     \end{subfigure}
        \caption{For the Irish politicians network with $a_{\nu}=5, b_{\nu}=10G$, where $G=10$: (a) the posterior distribution of the number of non-empty components $G_+$, and (b) the distribution of the effective latent space dimension $\hat{p}$.}
        \label{fig:irish_nu_mid}
\end{figure}

Supplementary results from applying the LSPCM to the Twitter networks are provided here, including the trace plots (Figures \ref{fig:football_trace}, \ref{fig:football_psi_trace}, \ref{fig:irish_trace} and \ref{fig:irish_psi_trace}), and the posterior distributions of the variance, shrinkage strength, and the $\alpha$ parameters (Figures \ref{fig:footballviolin} and  \ref{fig:irishviolin}). 
For the Irish politicians' network, Figures \ref{fig:irish_nu_high} and \ref{fig:irish_nu_mid} show the LSPCM clustering solution from a single MCMC chain at different values of $G$ and  $b_\nu$, which then informed the prior hyperparameter choices used in Section \ref{app:irish}. Figure \ref{fig:irish_cluster_bar_nu_high} and Table \ref{tab:irish_cross_high} show that under the hyperparameter settings detailed, the LSPCM estimated clustering solutions had high variance on the number of non-empty components and resulted in clusters with very small numbers, with some having as few as 3 nodes in a cluster. The context and knowledge of the political composition of the Irish political landscape prompted a lower value of $G$ and and a higher value of $b_\nu$ to be employed to better reflect the anticipated range of the likely numbers of clusters and to encourage stronger emptying of components to concentrate the nodes in fewer clusters for clearer insight to the clustering structure.

\clearpage
\begin{table}[hbt!]
\centering
\caption{For the Irish politicians network with $a_{\nu}=5, b_{\nu}=5G$, where $G=20$: Cross-tabulation of political party membership and the LSPCM representative cluster labels. 
}
\label{tab:irish_cross_high}
\scalebox{.85}{
\begin{tabular}{llccccccccccccccc}
\toprule
 && \multicolumn{15}{c}{\textbf{Cluster}}\\ 
& & \em 1  & \em 2  & \em 3   & \em 4  & \em 5 & \em 6 & \em 7 & \em 8 & \em 9 & \em 10 & \em 11 & \em 12 & \em 13 & \em 14 & \em 15 \\ \midrule
                   & \em Fine Gael            &   &   & 89 &   & 14 &   & 2 &   & 10 &   & 12 & 11 & 5 &   & \\
                   & \em Fianna Fáil          &   &   &    & 3 &   & 43  & 1 &   &   &   &   &   &  &  2 & \\
\textbf{Political} & \em Labour Party         & 71 &   &    &   &   &   & 2 &   &   & 6 &   &   &   &   & \\
\textbf{Parties}   & \em Independent          & 1  &   & 2  & 16 &   &   & 3 & 3 &   & 2 &   &   &   & 1 & 3 \\
                   & \em Green Party          &   &   &    & 5  &   &   &   & 2 &   &   &   &   &   &   & \\
                   & \em Sinn Féin            &   & 30 &    &   &   &   &   &   &   & 1 &   &   &   &   & \\
                   & \em United Left Alliance &   &   &    & 1  &   &   &   & 7 &   &   &   &   &   &   & \\

\bottomrule
\end{tabular}
}
\end{table}

\begin{table}[hbt!]
\centering
\caption{For the Irish politicians network with $a_{\nu}=5, b_{\nu}=10G$, where $G=10$: Cross-tabulation of political party membership and the LSPCM representative cluster labels. 
}
\label{tab:irish_cross_mid}
\begin{tabular}{llcccccccc}
\toprule
 && \multicolumn{8}{c}{\textbf{Cluster}}\\
& & \em 1  & \em 2  & \em 3   & \em 4  & \em 5 & \em 6 & \em 7 & \em 8  \\ \midrule
                   & \em Fine Gael            &   &   & 119 & 20  &  & & 2 & 2 \\
                   & \em Fianna Fáil          &   &   &    &  &  & 49  &  &   \\
\textbf{Political} & \em Labour Party         & 77 &   &    & 2  &  & & & \\
\textbf{Parties}   & \em Independent          & 6  &   & 2  & 2  & 16 &  & 5 & \\
                   & \em Green Party          & 1  &   &    &   & 6 & & & \\
                   & \em Sinn Féin            &   & 31 &    &   &  & & & \\
                   & \em United Left Alliance &   &   &    & 8  &  & & & \\

\bottomrule
\end{tabular}
\end{table}

\end{document}